\documentclass[12pt,preprint]{aastex}
\usepackage{a4wide,graphicx}

\newcommand{\degree}{$^\circ$}

\newcommand{\kms}{km~s$^{-1}$}
\newcommand{\ergcms}{erg~cm$^{-2}$~s$^{-1}$}
\newcommand{\magpersqa}{mag~arcsec$^{-2}$}
\newcommand{\sqa}{arcsec$^{2}$}
\newcommand{\sqam}{arcmin$^{2}$}

\newcommand{\othree}{[O~{\sc III}]}
\newcommand{\ntwo}{[N~{\sc II}]}

\newcommand{\Htwo}{H{\small II}}
\newcommand{\Halpha}{H{$\alpha$}}
\newcommand{\Lyalpha}{Ly-{$\alpha$}}
\newcommand{\ePNS}{PN.S}
\newcommand{\PN}{PN}
\newcommand{\PNe}{PN}

\newcommand{\FPN}{F_{\rm PN}}

\newcommand{\fback}{F_{\rm back}}
\newcommand{\SN}{{\rm S/N}}
\newcommand{\FWHM}{{\rm FWHM}}
\newcommand{\dpsf}{{\delta_{\rm psf}}}
\newcommand{\apsf}{{A_{\rm psf}}}
\newcommand{\ton}{{T_{\rm on}}}

\newcommand{\afibre}{{A_{\rm fibre}}}
\newcommand{\lmm}{l/mm}
\newcommand{\f}[1]{\rm f/{#1}}       
\newcommand{\PSF}{P}

\slugcomment{second revision for PASP \today}

\begin{document}
\title{The Planetary Nebulae Spectrograph:\\
the green light for Galaxy Kinematics}

\shorttitle{The PN.Spectrograph}
\shortauthors{Douglas et al.}

\author{N.G. Douglas}
\affil{Kapteyn Institute, Groningen, Netherlands}
\email{ndouglas@astro.rug.nl}

\author{M. Arnaboldi}
\affil{Osservatorio di Capodimonte, Naples, Italy}
\email{magda@na.astro.it}

\author{K.C. Freeman}
\affil{Research School of Astronomy \& Astrophysics, ANU, Canberra, Australia}
\email{kcf@mso.anu.edu.au}

\author{K. Kuijken}
\affil{Kapteyn Institute, Groningen, Netherlands}
\email{kuijken@astro.rug.nl}

\author{M. Merrifield}
\affil{School of Physics \& Astronomy, University of Nottingham, U.K.}
\email{michael.merrifield@nottingham.ac.uk}

\author{A. J. Romanowsky}
\affil{Kapteyn Institute, Groningen, Netherlands}
\email{romanow@astro.rug.nl}

\author{K. Taylor\altaffilmark{1}}
\affil{Anglo-Australian Observatory, Sydney, Australia} 
\email{kt@astro.caltech.edu}

\author{M. Capaccioli} 
\affil{Osservatorio di Capodimonte, Naples, Italy}
\email{capaccioli@astrna.na.astro.it}

\author{T. Axelrod} 
\affil{Research School of Astronomy \& Astrophysics, ANU, Canberra, Australia}
\email{tsa@merlin.anu.edu.au}

\author{R. Gilmozzi}
\affil{European Southern Observatory, Munich, Germany}
\email{rgilmozz@eso.org}

\author{J. Hart}
\affil{Research School of Astronomy \& Astrophysics, ANU, Canberra, Australia}
\email{john.hart@mso.anu.edu.au}

\author{G. Bloxham}
\affil{Research School of Astronomy \& Astrophysics, ANU, Canberra, Australia}
\email{gabe@mso.anu.edu.au}

\author{D. Jones}
\affil{Prime Optics, Eumundi, Australia}
\email{Prime\_Optics@bigpond.com}

\altaffiltext{1}{Currently at
California Institute of Technology, Pasadena, USA}

\begin{abstract}

Planetary nebulae are now well established as probes
of galaxy dynamics and as standard candles in distance
determinations.  Motivated by the need to improve the efficiency
of planetary nebulae searches and the speed with which their
radial velocities are determined, a dedicated instrument - the
Planetary Nebulae Spectrograph or \ePNS\ 
- has been designed and commissioned at the 4.2m
William Herschel Telescope.  The high optical efficiency of the
spectrograph results in the detection of typically $\sim 150$ \PNe\
in galaxies at the distance of the Virgo cluster in one night of
observations.  In the {\em same observation} the radial
velocities are obtained with an accuracy of $\sim 20$~\kms
\\
{\bf note that due to archival restrictions the figures have
been strongly compressed - please contact any of the authors for
a better preprint}
\end{abstract}

\keywords{planetary nebulae, radial velocity measurements, 
galactic dynamics, elliptical galaxies, photometry, 
astronomical instrumentation}

\section{Introduction}
\label{intro}

The study of the internal dynamics of galaxies provides some of the best
observational clues to their formation history and mass distribution,
including dark matter, but much of the interesting information is
inaccessible with conventional techniques.  Stellar kinematics, the most
important diagnostic, have generally been determined using
absorption-line spectra of the integrated light, the surface brightness
of which is such that only the inner parts of the galaxy can be observed
in a reasonable amount of telescope time.  In practical terms, it is
hard to measure the integrated stellar spectra beyond $1-2 R_e$ (the
``effective radius'', which contains half the galaxy's projected light). 
This is a serious limitation, because it is at larger radii that the
gravitation of the dark matter halo is likely to dominate, and that the
imprint of the galaxy's origins are likely to be found in its stellar
orbits. 

Alternative diagnostics of the gravitational potential include
measurement of the 21cm-wavelength emission from neutral hydrogen,
observations of \Htwo\ regions, and the observation of the
motions of globular clusters and planetary nebulae.  But neutral
hydrogen and \Htwo\ regions are effectively absent from early-
type galaxies, and globular clusters, not surprisingly,
have been found to be kinematically distinct from the stellar
population which is of primary interest. 

Planetary nebulae (\PNe) are part
of the post-main-sequence evolution of most stars with masses
in the range 0.8--8 M$_{\odot}$. 
Taking into account
the duration of the PN phase itself,
even in a galaxy with continuing star formation 
 most {\em observed} \PN\ 
will have progenitors in the range 1.5--2~M$_{\odot}$, corresponding
to a mean age of $\sim 1.5$~Gyr. In the case of early-type galaxies
the \PN\ are drawn from the same old population
that comprises most of the galaxy light.

\PNe\ are sufficiently bright to be detected in
quite distant galaxies and 
their radial velocities are readily
measured by a variety of techniques. 
Moreover, because they are easier to detect
at large galactocentric radius
where the background continuum is fainter, they 
represent the
crucially important complement to absorption line studies.
These properties make \PNe\ an ideal kinematic tracer for
the outer parts of such galaxies,
allowing the measurement of stellar kinematics
to be extended out to typically
$4-5 R_e$.

A common and well-tested technique for obtaining the radial velocities
of \PN\ consists of a narrow-band
imaging survey of [OIII] emission
to identify candidates, followed by a spectroscopic campaign to
obtain their spectra. This approach has in the past commonly
resulted in low yields (see \S\ref{mos}).

We have developed an alternative single-stage method,
which uses narrow-band slitless spectroscopy.  By obtaining two
sets of data with the spectra dispersed in different directions
(a technique we call ``counter-dispersed imaging'' or CDI),
one can identify
\PN\ {\em and} measure their radial velocities in a single
observation. The use of CDI by modifying existing
instrumentation has been so successful that we 
decided to design
and build a custom-made CDI spectrograph, with an
overall efficiency improvement of about a factor
of ten over a typical general-purpose spectrograph. 

The organisation of this paper
is as follows. In \S\ref{obs} we review
the important observational characteristics of \PN\
and discuss the implications of their luminosity
function for  the number of \PN\ which can be
detected.   In \S\ref{ot} we introduce 
CDI and compare it with
more traditional techniques. The PN.Spectrograph,
which is the main subject of this paper, is
presented in some detail in \S\ref{project}.
As well as a complete technical discussion,
and a short history of the project,
this section includes some of
the first images  obtained with the instrument, following
its recent successful commissioning. Further technical
information is presented in three appendices.

\section{Observation of extragalactic \PN}
\label{obs}

\begin{figure}
\includegraphics[width=12cm,angle=0]{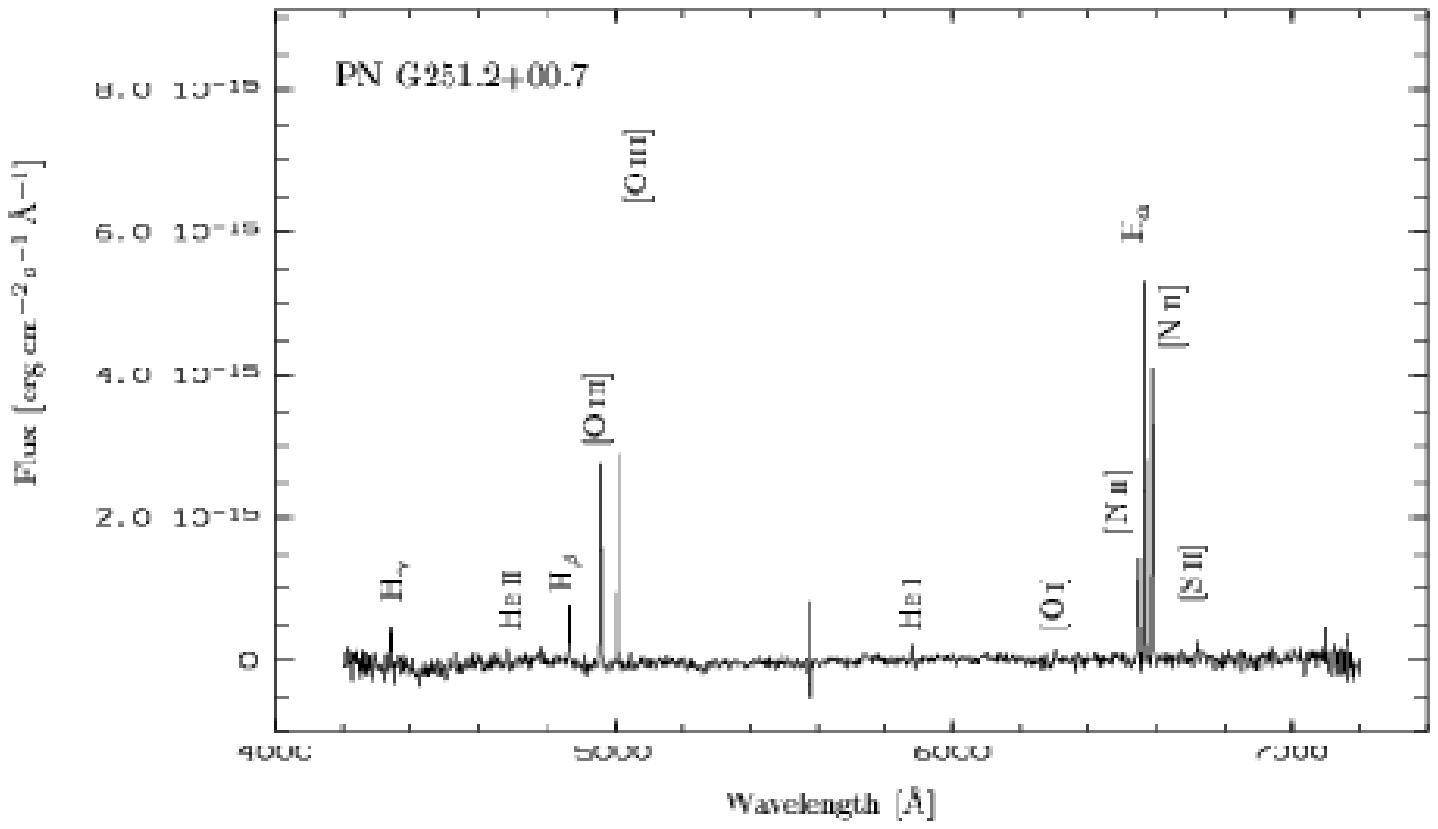}
\caption{A typical PN optical spectrum, showing the many emission
lines visible in such a nebula, as well as the dominance of the
\othree\ line. Figure kindly provided by F.~Kerber, ESO.}
\label{fig:pnspec}
\end{figure}

\PN, which are unresolved objects at
distances of 1~Mpc or more, can be detected
by means of their strong characteristic
emission lines (see
Fig.~\ref{fig:pnspec}).
The narrowness of the
lines (the intrinsic linewidth is usually less than 30\kms)
makes it relatively easy to measure the radial velocities of the
individual \PN.

The brightest line is usually
the 5007\AA\ line of \othree\ in which as much as 15\%\ of 
the central star's energy is emitted
\citep{djv92}.
The PN brightness in this line is conventionally given in magnitudes as 

 \begin{equation} 
  m_{5007} = -2.5 \log F_{5007} - 13.74
  \end{equation} 

where $ F_{5007}$ is in \ergcms\ and the constant is such that
a star of the same magnitude would appear to be equally bright
when observed through an ideal V-band filter. The \othree\ line
consists of a doublet 4959/5007\AA\ with the intensity ratio 1:3. Only
the brighter component of the doublet (rest wavelength in air
5006.843\AA) is included in the above magnitude. The absolute
magnitude $M_{5007}$ is defined in the usual way. 

\PN\ are found to have a well-behaved luminosity function  $\Phi$
which can be approximated by (dropping the {\sc 5007} subscript)

\begin{equation}
\Phi(M) \propto  e^{0.307M}[1-e^{3(M^*-M)}] 
\label{pnlf}
\end{equation}

where the parameter $M^*$, defining a sharp bright-end cutoff to
the luminosity function, was normalised using \PN\ in M31 to
$M^*$ = --4.48 \citep{Ciar89} and has since changed only
slightly \citep{Ciar02}.
The symbol    
$m^{*}$ is used for the apparent magnitude of the PNLF 
cutoff.

By fitting the above function to a
(complete) sample of \PN\ in a given galaxy a distance estimate
can be made.  This application of the PNLF, the characteristics
of which are under constant review (see e.g.  \citet{Ferrarese}),
will not concern us further here, but knowledge of the
shape of the PNLF is important in that it allows us to estimate the
relative number of \PN\ in a given system down to a certain
magnitude limit (see below). 

The absolute number of \PN\ present in a 
stellar system is usually characterised by
the number-density in the top dex
(2.5~magnitudes) of the PNLF, 
normalised to bolometric magnitude,
which takes a value 
\begin{equation}
\alpha_{2.5} \sim 7 - 50 \times (10^9 L_{\sun,{\rm bol}})^{-1} ,
\label{alpha25}
\end{equation}
depending on the colour \citep{Ciar91} and luminosity \citep{Ciar95}
of the galaxy. 
For practical purposes a number-density ($\beta_{2.5}$)
normalised to the B-band luminosity  is often used.

Eqn~\ref{pnlf} can be used to compute the cumulative luminosity
function.  The results, normalised to $\alpha_{2.5}$, are shown in 
Table~\ref{table:PNLFsummed}.  This shows for example that to double the
number of \PN\ detected in a given galaxy with respect to a complete
sample down to $M-M^* = 2.5$ one needs to reach a limit nearly 1.5
magnitudes, or a factor 4 in flux, fainter. 

As a numerical example we consider the Virgo galaxy NGC~4374, with an
assumed distance of 15 Mpc and a foreground extinction of 0.2 mag.  The
brightest \PN\ in this galaxy would be expected to have an apparent
magnitude of $m^{*}_{5007} \simeq 26.6$, corresponding to an \othree\ flux
of $7.3 \times 10^{-17}$ \ergcms.  This converts to a detection rate in
the \othree\ line of 0.7 photon per second for a 4m telescope with 30\%
system efficiency. For NGC 4374, $\alpha_{2.5} = 16.7$ \citep{jcf90}, 
and with the appropriate bolometric correction (BC = $m_{\rm bol}-m_V$ =  -0.83)
and colour index (B-V = 0.97) 
this corresponds to $\beta_{2.5} = 47.7$. Correcting the apparent magnitude
($m_{\rm B} = 10.3$) for distance and extinction gives
$M_{\rm B} = -20.8$, corresponding to $32.5 \times 10^9 L_{\sun,{\rm B}}$.
Thus one would expect $\sim 1550$ \PN\ in the top dex of the
luminosity function or, using Table~\ref{table:PNLFsummed}, $\sim 710$
down to the magnitude limit  $m^{*}_{5007} = 28.1$.

By definition half of these would fall within one
effective radius of the centre, where the noise contribution from the 
galaxy light is very significant, and be hard to detect. For NGC~4374
the B-band background light is 22.2 \magpersqa\ at $1 R_e$, which
corresponds to 50\% of the dark night sky background at a good
observing site.

To put the above number of \PN\ in perspective, note that 
$\sim$1000 radial velocities are necessary to 
nonparametrically constrain the dynamical properties of a hot
stellar system (\citet{MerSa93}, \citet{MerTr93}).
However, early-type galaxies normally have integrated light
kinematical measurements available to $\sim 1 R_e$, the 
addition of which dramatically decreases the number of discrete velocity
measurements needed.  In such cases, $\sim$100--200 velocities
can place strong constraints on a galaxy's mass distribution
(\citet{Sag00}, \citet{RoKo01}).
If 200 PN velocities are measured outside 1~$R_e$, $\sim$50
will be outside 3~$R_e$, a sample that can reliably constrain
the rotational properties of the galaxy's outer parts
(\citet{Na01}).


\clearpage
\begin{table}
\begin{center}
\caption{The cumulative PN luminosity function (top row) 
as a function of magnitude (bottom row).\label{table:PNLFsummed}}
 \begin{tabular}{|r|r|r|r|r|r|r|r|r|r|r|r|}
 \tableline\tableline
          & 0.00 & 0.08 & 0.24 & 0.46 & 0.71 & 1.00 & 1.34 & 1.74 & 2.21 & 2.75 &3.39\\
$M - M^*$ & 0.0  & 0.5  & 1.0  & 1.5  & 2.0  &  2.5 & 3.0  & 3.5  & 4.0 & 4.5 & 5.0 \\
\tableline
\end{tabular}
\end{center} 
\end{table}

\section{Observing techniques}
\label{ot}

To study the kinematics of extragalactic \PN, there are two
aspects:  finding the \PN\ and obtaining their radial velocities.
In the following sections we
discuss the techniques applied to this problem so far. Simple formulae
are given for the  
\SN\ achieved with each technique, and then used to compute the integration
time required to reach $S/N = 10$ for a PN with given magnitude. 
This is a reasonable
benchmark since experiments show that detection completeness
varies from nearly 100\% at 
$S/N \ge 9$ to 0\% at $S/N = 4$  \citep{Ciar87}. 
We compute the noise as the quadratic sum of poisson noise
from the object, sky background, and detector read out.
We adopt the notation
of Table~\ref{table:notation} and the default values
therein, unless otherwise specified. 

The ``system efficiency'' which is quoted for multi-object spectroscopy
(MOS) and other techniques is based on typical limiting magnitudes
actually achieved with current instrumentation at 4-m class telescopes. 
By definition it includes telescope, filter and detector efficiency as
well as the efficiency of the instrument proper, but it should be
stressed that light loss due to the finite size of the fibre or slit,
and hence also astrometric, pointing, and tracking errors affect the
performance of MOS instruments.  This can lead to values of the system
efficiency which are smaller by a factor of two or more 
than the nominal optimum instrument performance.

In these and other calculations in the paper we include atmospheric
losses of 0.15~mag corresponding to an airmass of 1.0 at a good
observing site. 

\clearpage
\begin{table}
\begin{center}
\caption{Parameters used in the computation of 
integration time required for a given \SN. \label{table:notation}}
\begin{tabular}{|l|l|}
\tableline\tableline
\multicolumn{1}{c}{Symbol}&\multicolumn{1}{c}{Description}\\
\tableline
$S/N$   & final signal-to-noise ratio for the detection of a PN \\         
	& against a background, including all
		noise contributions\\
FWHM    & the FWHM of the seeing point-spread-function (default 1\arcsec)\\
$\apsf$ & in imaging, the area\tablenotemark{a} \,  $\pi (\frac{2}{3} \FWHM)^2$ over which the $S/N$ is calculated \\
$\FPN$  & detected \othree\ flux from the object in photons/sec\\
$T$     & total integration time, in seconds\\
$\ton$  & integration time spent on-band, in seconds\\
$\fback$& total detected background flux\tablenotemark{b} \,
	       in photons/sec/\AA/\sqa\\
$B$     & bandwidth of the \othree\ band-pass filter (default 30.0\AA)\\
$\afibre$ & in a fibre-fed spectrograph, the effective size of the fibre,\\
         & as projected onto the sky (default 1.5 \sqa) \\  
$\Delta_\lambda$     & in a spectrograph, width of a
resolution element (default 3.0\AA) \\ 
$n_p$     & in imaging, number of pixels over which the integration is calculated \\
	& (default $\apsf * 9$)\\
$n_p$   & in spectroscopy, number of pixels over which the integration is calculated \\
	& (default 9)\\
$n_r$   & number of times the CCD is readout per integration \\
	& (default 2 per hour)\\
$r$     & read-noise, converted to photons (default 4.0)\\ 
\tableline
\end{tabular}
\tablenotetext{a}{See \citet{naylor98}}
\tablenotetext{b}{This is calculated from 21.0 \magpersqa,  
being 21.4 \magpersqa\ for sky and 22.2  \magpersqa\
for the surface brightness of a  typical galaxy at 1~$R_e$}
\end{center}  
\end{table}

\subsection{Survey/multi-object spectroscopy}
\label{mos}
Most of the \PN\ work on external galaxies has been done by first
performing an imaging survey to obtain a sample of \PN, and then
following this up with multi-fibre or multi-slit spectroscopy of
the individual objects to obtain their radial velocities  \citep{Hui95}.

Detection of the \PN\ in \othree\ emission is usually done by
comparing images made with on-band and off-band filters
\citep{jcf90}, which allows PN candidates to be distinguished from
foreground stars.  For maximum sensitivity, the on-band filter
should only be wide enough to accommodate the expected range of
velocities, while the off-band filter can be broader.  The
positions of the \PN\ must be determined to at least
0.5~arcsec precision in order to set up a subsequent observing
campaign with standard
multi-object spectroscopy (MOS).

For the detection of a PN in a narrow-band survey 

\begin{equation}
 S/N = \FPN \ton / 
 \sqrt{\FPN \ton + \apsf \fback  B \ton + n_p n_r r^2}
\label{sn:nb}
\end{equation}

where the symbols are defined in Table~\ref{table:notation}.  The
assumption is made that the off-band integration time ($T - \ton$) is
chosen such as to reach the {\em same level of sky background} (owing to
the wider bandpass the image can be obtained relatively quickly).  For
useful S/N values, the read-noise is usually negligible.

As an illustration, consider the case of a survey carried out at
a telescope of 4.0m diameter. The overall detection efficiency
is assumed to be 0.4, and $T$ is divided between narrow-band and
broad-band imaging in the ratio 80:20 ($\ton = 0.8\times T$). 
The assumed background (see Table~\ref{table:notation}) 
corresponds  to 
 $\fback = 0.20$~photons/sec/\AA/\sqa\ for this case. The goal is
to detect objects down to $ m_{5007} = 28$ with $\SN~\sim 10$.
Equation~\ref{sn:nb} with default parameters indicates that
this can be achieved in $T = 6.2$~hours. 

The spectroscopy can be done efficiently with fibre or multislit 
spectroscopy, and the detection of the \othree\ line is 
usually dominated by photon noise. 
For example, for fibre spectroscopy at spectral resolution $\Delta_\lambda$,
this being the bandwidth corresponding to the projected width of
the fibre in the spectra,
\begin{equation}
 S/N = \FPN T / 
\sqrt{\FPN T + \afibre \fback \Delta_\lambda T +  n_p n_r r^2}
\end{equation}

For this case, the system efficiency measured in practice (see
\S~\ref{ot}) is usually in the range 0.05-0.1, and here we adopt 0.1. 
With similar observing parameters to the previous example, and default
values for $\Delta_\lambda$ and $\afibre$, the multi-object spectrograph
should then return spectra with \SN~$\approx 10$ for a $ m_{5007} = 28$
object in about $T = 3.3$~hours. 

Since several fibre configurations might be required to obtain
all of the spectra (as a result of limitations on the number
of fibres or on their positioning) we
must multiply $T$ by some factor, typically $\sim 2 - 4$. Total
time for the project, even with these ideal assumptions, 
is thus $\sim 13 - 19$~hours. 
However, as discussed in \S\ref{aCDI},
several practical
problems arise  in the MOS follow-up to narrow-band imaging,
significantly reducing the yield of \PNe\ velocities obtained.

\subsection{Fabry-Perot interferometry}

A Fabry-Perot system produces a cube of narrow-band images and
is in principle well suited to observing extragalactic \PN. Its
large field of view (10\arcmin\ at a 4-m telescope is typical)
can encompass typical galaxies out to five effective radii at the
distance of the Virgo  Cluster.  
\citet{Trem95} used this technique to study the \PN\ in the SB0
galaxy NGC~3384.

The theoretical signal-to-noise for the
detection of an emission-line object in the appropriate step from a
wavelength scan in which a total bandpass  $B$ is scanned over $N$
resolution elements is
 
\begin{equation}
 S/N = (\FPN T/ N) / 
\sqrt{(\FPN T/ N) + \apsf \fback\  (B/N)  (T/N) + 2 n_p r^2}
\label{sn:fp}
\end{equation}

where now $n_r = 1$ is assumed since the integration time per 
step is usually short. The factor 2 in the read-noise term allows
for the fact that the actual step size needs to be such that
there are two steps per resolution element.

When sky noise dominates, the Fabry-Perot procedure is
equivalent to observing each planetary nebula by direct detection
through an ideal filter (of bandwidth $B/N$) for time $T/N$, and the
corresponding $S/N$ is similar to that of narrow-band imaging
(with bandwidth $B$ and time $T$). However both read out noise and
the poisson noise of the object being detected reduce the $S/N$
significantly. Neverthless, since  the Fabry-Perot
requires no follow-up spectroscopy, it is competitive with
other techniques.

With other parameters as before we now assume
an overall detection efficiency of 0.3 including the Fabry-
Perot etalon, and $N = 40$,  giving a
nominal resolution element of about 45~\kms. 
Equation~\ref{sn:fp} then shows that 21.6~hours observing is
required to cover the passband and to reach $S/N \sim 10$ in each
resolution element for the same magnitude limit as before.
In the absence of read-noise, the same limit would be reached in
about 13~hours.

\subsection{Counter-dispersed imaging (CDI)}

\label{cdi}

\begin{figure}

\includegraphics[width=12cm,angle=0]{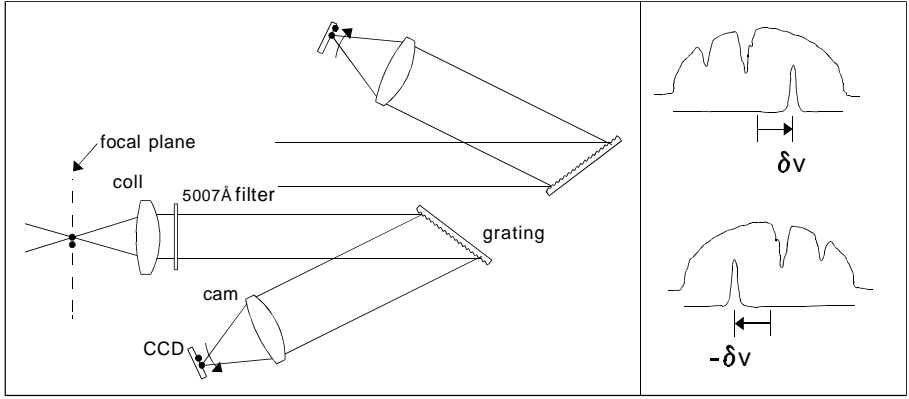}
\caption{Left panel: illustration of counter-dispersed imaging -
the arrow denoting the dispersion direction. Right panel: (left) 
the corresponding
images of a star and planetary nebula shown schematically. 
}
\label{fig:pnseek1e}
\end{figure}

CDI, which  
has antecedents in the work of Fehrenbach (e.g.  \citet{Fehr47}),
is illustrated schematically in Fig.~\ref{fig:pnseek1e}. The left panel
shows that two
images of a given field must be taken through a slitless
spectrograph equipped with a \othree\ filter.  The field is identical in
the two cases but the dispersion direction is reversed. This can be
done simply by rotating the spectrograph sequentially
between position angles differing by 180\degree.

As shown in the right-hand panel of the Figure, stars in the field
appear in the images as short segments of spectra whose extent depends
upon the dispersion of the grating and on the filter bandwidth.  The
\PN, on the other hand, are detected only on the basis of their single
emission line and thus appear as point sources.  The exact position at
which each PN is detected in the slitless images depends upon the
actual position of the PN on the sky and its precise wavelength. 
As the two images are counter-dispersed it can readily be seen that, 
after matching up
the detected \PN\ in pairs in the final images, one directly obtains
relative velocities and positions.  The velocities can be put on an
absolute scale if the slitless observations include an arc line
calibration through a slit. This discussion of the velocity solution
is somewhat oversimplified, and we
return to this topic later (\S\ref{calibration})
in more detail.

For a detected PN the S/N in each image takes the form

\begin{equation}
S/N = (\FPN  \ton) / \sqrt{\FPN  \ton + \apsf \fback  B  \ton + n_r
n_p  r^2} \label{sn:cdi}
\end{equation}

where the total observing time $T$ is split between the two
position angles ($\ton = T/2$).  
With the same parameters as before except for an 
overall system efficiency of 0.3 (somewhat
less than for narrow-band imaging because of the presence of a
grating in the beam), each CDI
image achieves $\SN
\approx$ 10 in $T = 13.1$~hours. In the case of
simultaneous CDI (\S\ref{dc}) 
there will be twice as many images and therefore
more read-noise, but this has negligible effect.

\subsection{Advantages of CDI}
\label{aCDI}

\begin{figure}
\includegraphics[width=12cm,angle=0]{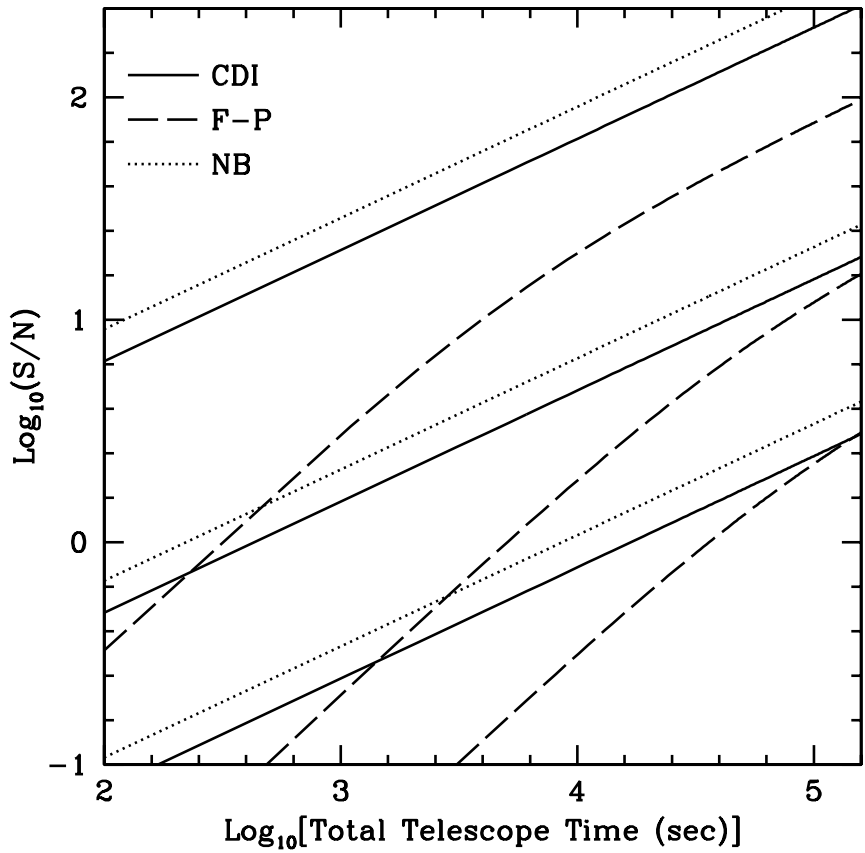}
\caption{
Attained signal to noise ratio as function of telescope time, for
three different techniques of searching for planetary nebulae.
Solid line:
counter-dispersed imaging (\ePNS); dashed line: Fabry-Perot;
dotted line: narrow-band imaging (not including
spectroscopic follow-up).  For each, three curves are
plotted, referring (top to bottom) to $m_{5007}=25$, 28, and 30. 
  }
\label{fig:sn}
\end{figure}

The results of the previous sections are summarised
in Fig.~\ref{fig:sn}, which shows the \SN\ reached
with a 4.0m diameter telescope
by different techniques - CDI, direct imaging (not
including the follow-up spectroscopy), and Fabry-Perot
spectroscopy - as a function of integration time. 
The same parameters were used as before.

As can be seen, the FP technique is inferior except for very faint
sources, being more sensitive to read noise on the one hand, and on the
other hand not as
efficient as the other techniques for the bright sources which are not
background limited.

Narrow-band imaging is somewhat more efficient than CDI,
but the former still requires the spectroscopic
follow-up work.  Multi-object spectroscopy can in principle be very
efficient, but the need for several MOS configurations alone will
usually mean that a campaign based on surveying and follow-up
spectroscopy will require more total telescope time for a given yield
than would be required with CDI.

Moreover there are additional problems.
The imaging and the spectroscopy are rarely
carried out on the same instrument, or even the same telescope.
Observing time must be obtained twice, possibly at different
telescopes and usually in different seasons, with greater risk
of poor weather and incomplete data sets.  
Astrometric errors in transferring from the
survey position to the fibre position sometimes
mean that the spectrum is not obtained (e.g.  \citet{Arna96} 
recovered only 19 PN spectra
from a survey set of 141 in NGC~4406), and the yield is further
reduced by the fact, recently discovered (\citet{Ku00}, \citet{Ar02}),
that the sample from the narrow-band imaging technique is
sensitive to contamination from misidentified
continuum objects.

CDI can be done with one instrument and at one telescope, 
and although still
vulnerable to the possibility of incomplete or inhomogeneous datasets
as a result of changes in observing conditions,  this 
problem can be eliminated by obtaining the CDI images simultaneously.

In CDI the relative velocities are extremely accurate, being determined
by the centroiding of images on the stable medium of a CCD.  The radial
velocity precision is comparable to that of fibre spectrographs of
similar spectral resolution.  We note that multi-slit velocities are
typically less precise: they are subject to significant errors resulting
from object positioning within the slit. 

Accurate photometry requires, amongst other things, knowledge of the
filter profile. Generally manufacturer's curves are generated with
very high focal ratio beams and at laboratory temperatures.
In CDI the filter profile can be determined from the dispersed images of
foreground stars, at the focal ratio and temperature appropriate to the
observations and as a function of position on the detector, so accurate
photometry is possible. 
 It is more difficult to determine the filter
characteristics in direct imaging mode.  Note also that with the
assumptions made earlier, that a given object is
detected in {\em each} image
with the same value of \SN, the effective 
integration time on which PN fluxes are based in CDI is twice
that of the narrow-band image.

One disadvantage of CDI, relative to more conventional
spectroscopic techniques, is its limited ability to 
distinguish between
\PN\ and other emission-line objects like unresolved \Htwo\
regions, Ly$\alpha$  emitters at redshift $z\sim 3.1$ and
starburst galaxies at $z\sim 0.3$. 
However on average the
number of expected high-$z$ contaminants in a field of our size 
(100 \sqam) is only 2-3 \citep{cfk02}, and these can usually be recognised
by their large linewidth, 
which the \ePNS\ has some ability to resolve.  
Further identification of these objects,
and \Htwo\ regions (rare in early-type galaxies) 
could be accomplished with an Halpha arm (see \S\ref{udha}).
 
\subsection{Results with CDI}
\label{rCDI}

The CDI idea was systematically 
investigated in 1994-95 by Douglas and Taylor 
(at that time both at the Anglo-Australian Observatory (AAO)) and
Freeman. The first published
experiments, confirming with high accuracy positions and velocities
for \PN\ which had previously been detected in Cen~A, 
were carried out by removing the slit
unit from the RGO spectrograph at the 3.9~m AAT  \citep{DT99}. 
Subsequently the ISIS spectrograph at the 4.2~m WHT was used
between 1997 and 2000
to obtain the kinematics of M94 \citep{isis} and
several S0 galaxies, and in 1998 an attempt was  
made to reach planetaries in the E5 galaxy NGC~1344
using CDI at the ESO 3.6~m.
 In these cases the counter-dispersed
images were obtained by rotating the spectrograph.  
We note the related technique used by \citet{mendez01}, who used
an undispersed narrow-band image and a single slitless dispersed
image to detect extragalactic \PNe\ and to measure their velocities. 

The principal gain in designing a {\em dedicated} instrument
is the possibility of increasing both the optical efficiency and
field size. The use of such an instrument can enable 
projects to be carried out at a 4m telescope which would
otherwise require an 8m telescope with general-purpose 
instrumentation. 

\section{The PN.Spectrograph project}
\label{project}

These considerations led us to consider the construction of 
a purpose-built instrument.  In this section we present
first the design philosophy, then 
optical and mechanical details, before returning to the issue
of calibration. A chronology of the  PN.Spectrograph project
is given, and the first observational results are presented.

\subsection{Design considerations}
\label{dc}
An important and unique driver of our \ePNS\ design was that it should
deliver {\em simultaneous} counter-dispersed imaging, avoiding the need
to rotate the instrument between the two position angles.  
As described later in this section,  this was achieved
by using a matched pair of gratings and  cameras, creating what is
in effect two spectrographs back-to-back\footnote{One may ask whether two is in fact the optimum number of
images to use. For example an undispersed third image
could easily be generated by replacing the grating(s) with a mirror
for part of the observing time, and a third detection
would seem to
provide a much more robust rejection of false identifications
arising because of noise peaks. We investigated this, and other
performance characteristics, with simulated images, and found that
varying the distribution of observing time between the three arms
can easily give worse performance than for the two-arm
case, but apparently never better.}.
The increased
complexity  is compensated by several advantages: the CDI
data sets are always matched in quality and integration time, and the
velocity determination is more accurate since all calibrations are done
at the same time.  Differential flexure is less and can be more easily
monitored.

%
%

The original \ePNS\ design studies were for an \f{15}\ instrument to operate at
the Anglo-Australian Telescope (3.9m aperture) and/or the ESO VLT (8m). 
However, driven by considerations of cost and telescope access, we
decided instead to design the instrument to operate at both the UK/NL
William Herschel Telescope (4.2m) and the Italian Telescopio Nazionale
Galileo (3.5m), both located on the island of La Palma.  The TNG is
based on the ESO NTT and has two \f{11} Nasmyth focal stations with
instrument derotators.  
The WHT also has a Nasmyth platform
for interchangeable instrumentation but there the field of view is
rather restricted owing to the use of an optical derotator.  We
therefore opted for the WHT \f{11} Cassegrain focus with instrument
derotator.  This station also includes a suite of calibration lamps. 
Consistent with this choice of telescopes and focal stations, the
field-of-view was chosen to be $\sim 10$\arcmin\ square.  To be able to
optimize the throughput as much as possible, we designed for the
smallest wavelength range consistent with the range of galaxy systemic
velocities likely to be of interest ($\pm 1800$~\kms). 

The most critical
component in maximising the instrument efficiency over a narrow
wavelength range is the grating.  The performance of standard commercial
blazed gratings were examined with the assistance of 
computer codes\footnote{Optimization studies were carried out using the
code {\tt gtgr4}, developed by T.C.  McPhedran at the University of
Sydney and L.C.  Botten at the University of Technology, Sydney.  This
gave good results for gratings of infinite conductivity.  Later we used
results kindly provided by Daniel Maystre using code based on his
integral theory, which allows for the finite conductivity of the
aluminium layer \citep{may78}.}.

For the CDI detection of \PN\ the dispersion should not be too high
because of three effects: firstly, field is lost since \PN\ whose images
are dispersed outside {\em either} of the counter-dispersed frames
cannot be used, secondly the dispersed images of foreground stars
obliterate a fraction of the field, and thirdly the increased separation
between CDI pairs increases the degree of confusion, making it more
likely that some objects cannot be unambiguously paired up.  On the
other hand, the dispersion should be high enough for the required
velocity precision. Calculations show that it should be possible
to centroid a well-sampled PSF with given $S/N$
to an accuracy of $\dpsf \sim \frac{0.7}{S/N}\FWHM$ (see Appendix~\ref{centr}).
Given a detection (in each arm)
with $S/N = 5$ it follows that the {\em separation} of the two centroids
can be determined with an accuracy of 
$\sqrt{2} \dpsf \sim 0.2~$\FWHM. 
Assuming a 
required velocity precision of $\sim 20$~\kms, which is less than 
the internal velocity width of the \PN\ emission lines, the dispersion
should be chosen such that the \FWHM\
corresponds to $200$~\kms\ or 3.3\AA\
(recall that the dispersion is effectively {\em doubled} in CDI
-- see Fig.~\ref{fig:pnseek1e}).
For \FWHM\ $ \sim 3$~pixels it follows that the dispersion 
should be about 1.0~pixel/\AA.  For a filter of 
width 100\AA\ at the 10\% level (\FWHM~$\sim 35$\AA) the stellar trails will 
extend over an acceptable $\sim 100$ pixels.

\begin{figure}

\includegraphics[width=15cm,angle=0]{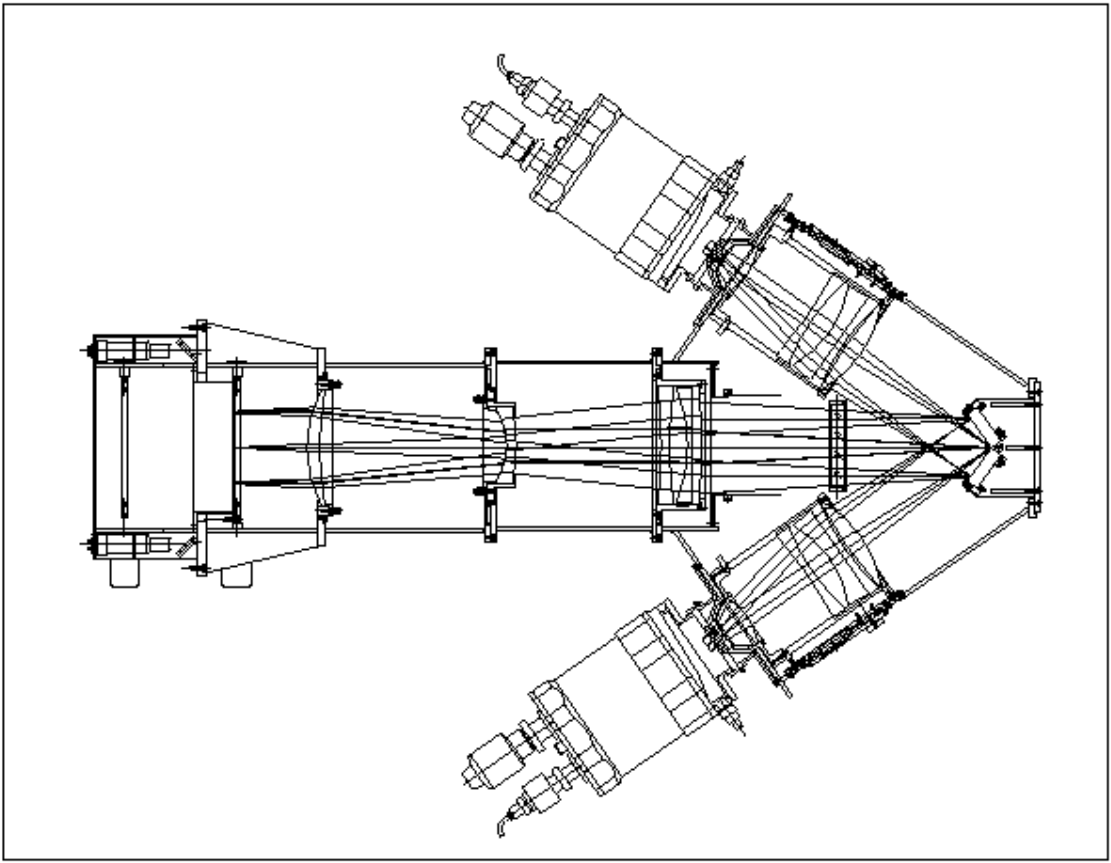} 
 \caption{\ePNS\ schematic. 
The pupil-splitting dual
grating mount is at the right of the figure, with
the interference filter just to the left.
The camera optics appear to be 
oversized and laterally offset, but
these are  artefacts due to the
projection chosen and to the fact that the pupil is
split. 
}

 \label{fig:ePNS} 
 \end{figure}

\subsection{Optics} 
\label{optics}

The optical system of the \ePNS\ (see Fig.~\ref{fig:ePNS}) consists of a collimator
with 5 elements in 4 groups, the largest lens being 240~mm in diameter,
a  narrow-band filter, twin 600 \lmm\ gratings, and a pair of nominally
identical cameras (6 elements in 5 groups) in which the largest element
has 210~mm diameter.    For ease of manufacture the design was restricted
to spherical surfaces, despite the fact that the anamorphic reflection at the 
gratings (transforming a nominally
square field into one with with an aspect ratio of 0.9) presented a
significant challenge. 

\begin{figure}

\includegraphics[width=12cm,angle=0]{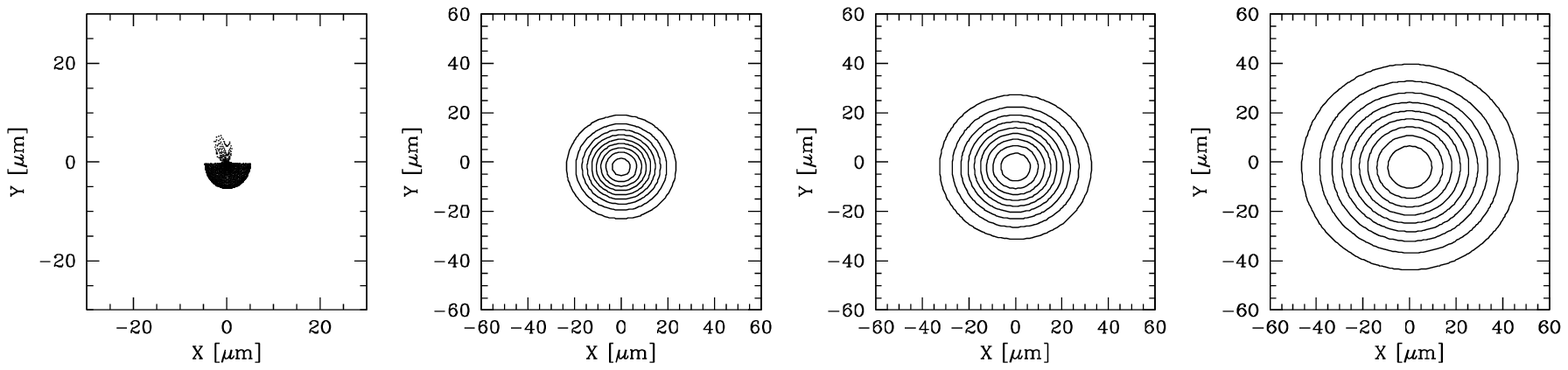}
\includegraphics[width=12cm,angle=0]{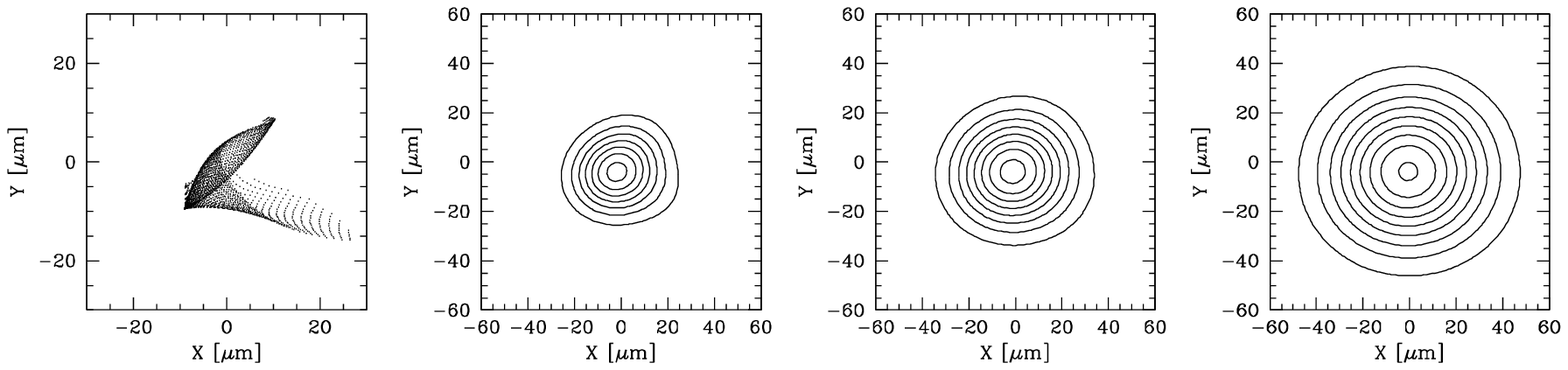}
\includegraphics[width=12cm,angle=0]{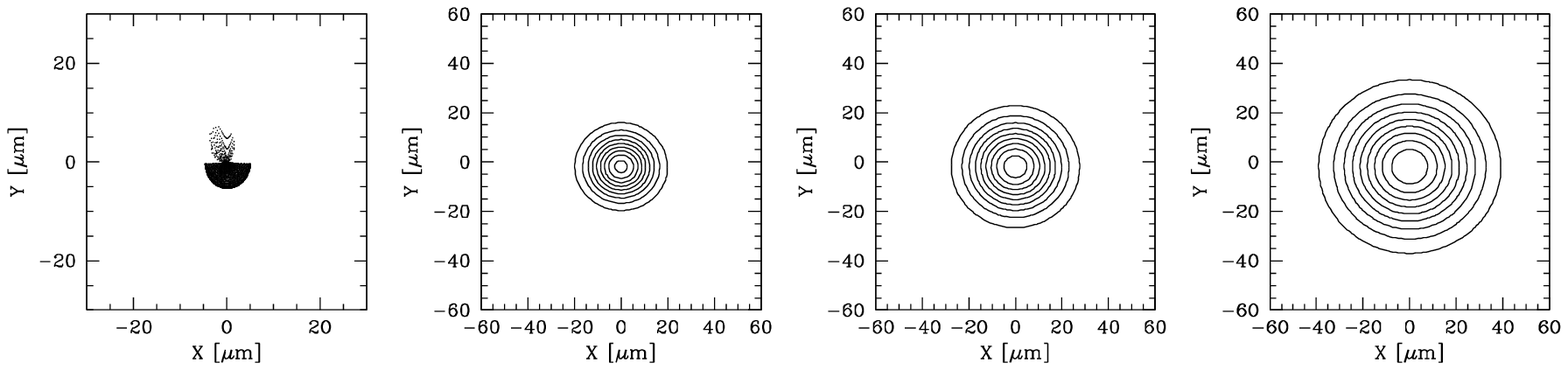}
\includegraphics[width=12cm,angle=0]{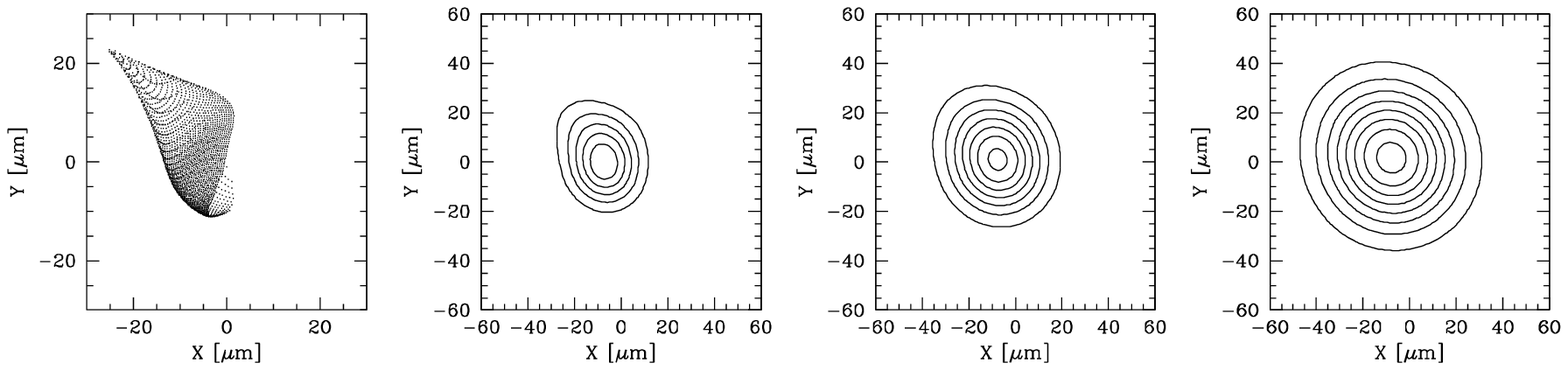}

 \caption{Spot diagrams for one arm of the spectrograph, at 5010\AA.
The left column shows the raytraced spot diagram for (from top
to bottom)  the WHT at field centre, the WHT in a corner of the field,
the TNG at field centre, and the TNG in a corner of the field (the
box is 60\micron\ square) - note the very asymmetric spot
pattern due to the half-pupil.
Subsequent columns show the same spots convolved (from left to right)
with gaussian seeing of 0.5, 0.7 and 1.0 arcsec, where 
the box size is now 120\micron\ square and allowance
has been made for the anamorphic ratio of 0.9, which compresses
the images in the vertical (dispersion) direction.}

\label{fig:spots}
\end{figure}

Table~\ref{table:oparameters} lists the optical parameters of the instrument
as built. The indicated operating range is nominal -- chromatic effects are not severe and the
instrument will function slightly outside this band, though the
coatings will not be optimal. The computed imagery  (Fig.~\ref{fig:spots})
is valid over this range.

The gratings are standard items except
that we requested 140 x 80 x 24 mm substrates replicated to the
very edge, and bevelled at the grating input angle so as 
to facilitate butting with
minimal `dead' area. We required tight tolerance on the alignment
of the grooves with the substrate edges so as to keep the dispersion
directions in the two cameras parallel.

The grating input angle between the collimator axis and the grating
normal is such that at a wavelength of 5010\AA\ the centre of the field
is centered on the CCD.  The gratings are in principle at a fixed angle,
and the detectors at a fixed position, so that as the wavelength is
changed the centre of the field moves laterally across the chip.  As the
imaging quality of the spectrograph is rather closely matched to the
size of the chip, it may be said that the usable field gradually ``walks
off'' the detector.  The ``walkoff rate'' is about the same as the
dispersion, i.e.  17.4~\micron\ per \AA, or $\pm 1.04$~mm over the full
operating range of the instrument ($\pm 60$\AA).  This is only 3.8\% of
the field, however, so although the grating tilts could be manually
adjusted to recentre the image it is not anticipated that this will be
done regularly.

\clearpage
\begin{table}
\begin{center}
\caption{Final parameters of spectrograph optics.\label{table:oparameters}}
\begin{tabular}{|l|l|}
\tableline\tableline
        Operating range & 4950-5070\AA\ \\  

      Grating(s)      & Spectronic\tablenotemark{a}\ 35-53 600 8.5\degree\ blaze \\

        Input angle (at grating)     & 26.6\degree \\
        Output angle (in first order)     & 8.4\degree (on-axis at 5010\AA)\\
        I.A.\tablenotemark{b}           & 35.0\degree \\   

     Collimator     & F.L. = 1291 mm\\
     Camera         & F.L. = 287 mm\\

        Image quality         & EER\tablenotemark{c} \, $\leq 18$~\micron \\
     Dispersion  & 0.0174 mm/\AA   \\
\tableline
\end{tabular}
\tablenotetext{a}{formerly Milton-Roy}
\tablenotetext{b}{I.A. (included angle) is the fixed value
of the angle betwen the optical axes of the collimator and camera}
 \tablenotetext{c}{EER (enclosed energy radius) is the (worst case)
radius of the smallest circle enclosing a certain percentage of the
light, here 90\% - the 
value given applies for use at both the WHT and the TNG}
\end{center} 
\end{table}

The design goal was for a total system efficiency
of at least 30\% at the WHT, about twice that of the general-purpose
spectrograph (ISIS) at that telescope, with $\sim 5$ times larger field.
The estimates in Table~\ref{table:calceff}, based on the
final optical design, published data and calculations, suggested that
this was feasible.

\clearpage
\begin{table}
\begin{center}
\caption{Estimated Optical Efficiencies.\label{table:calceff}}
\begin{tabular}{|l|l|l|}
\tableline\tableline
\multicolumn{1}{c}{Component}	& 
\multicolumn{1}{c}{Efficiency}	& 
\multicolumn{1}{c}{Basis of estimate} \\
\tableline
Grating 	& 0.83  & E.M. calculation assuming aluminium at 5010\AA \\
Lenses  	& 0.97& reflection loss (0.1\% at 10, and 
                                 0.2\% at 8 air/glass interfaces)\\
Lenses  	& 0.95 & bulk absorption, total path length 213~mm \\  
Window  	& 0.98 & dewar window without special coating   \\
&&\\
\ePNS\ total    & 0.73 & \\
Filter  	& 0.80 & manufacturer's specifications \\
Detector	& 0.85/0.90 & ING/TNG measurements \\
Telescope	& 0.72/0.63 & observatory calculations for WHT Cassegrain/TNG Nasymyth \\
&& \\
System		& 0.37/0.33 & at WHT/TNG\\
&& \\
\tableline
\end{tabular}
\end{center} 
\end{table}

Although the two telescopes for which the
instrument has been designed (\S\ref{dc})
are of similar diameter and have the same
focal ratio, the optics are sufficiently different in detail that it was
difficult to produce a single \ePNS\ design for both.  Full control of
the aberrations appeared to require interchangeable optical elements,
which introduced too much cost and complexity.  We agreed on a single
design in which the imaging quality is compromised, but not noticeably
so under normal observing conditions (see Fig.~\ref{fig:spots}).

Parameters which depend on the telescope are listed in Table~\ref{at:telescope}.
In this table a pixel size of 13.5~\micron\ has been assumed,  in agreement
with the detectors used for commissioning.  The images
are slightly better sampled at the WHT owing to the longer focal length,
but with this pixel size 
would still be adequately sampled at the TNG down to 0.7\arcsec\ seeing.
The field sizes indicated  correspond to the projected extent of the 
2048~x~2048 element detector currently in use. At both telescopes
the unvignetted field is about 10\arcmin\ square.

\clearpage
\begin{table}
\begin{center}
\caption{Parameters of 
	the PN.Spectrograph at Telescopes.\label{at:telescope}}
\begin{tabular}{|l|l|l|} 
\tableline\tableline
     Telescope &  4.2m WHT & 3.5m TNG\\ 
\tableline\tableline
     Focal station                   & f/10.942 Cassegrain &  f/10.755  Nasmyth\\ 
Diameter of pupil (at grating)       &    117.9 mm          & 120.0 mm\\
     Scale (pxl/arcsec)      & & \\
     ~~~~(in dispersed direction)  & 3.32 & 2.72 \\
        ~~~~(in spatial direction)      & 3.67 & 3.01\\
     Nominal field\tablenotemark{a} \, (arcmin)        & 11.43 x 10.34        &  13.95 x 12.59\\
\tableline
\end{tabular}
\end{center} 
\tablenotetext{a}{see text}
\end{table}

Details of the optical design (ZEMAX files) ``as built'' can be obtained
from the project website\footnote{www.astro.rug.nl/$\sim$pns} 
or on request to the team.  In addition,
optimised designs for the WHT and for the TNG are also available. 

The  V-type coatings on the camera optics (10 surfaces in each camera)
were 
specified as having maximum reflectance of 0.1\% within the operating range.
For the collimator (8 surfaces) a W-type coating was specified with
a  maximum reflectance of 0.2\% over the operating range
and 0.3\% over a secondary range covering 6488  - 6645\AA.
This was to allow the possibility of the addition of a
\Halpha-camera 
(\S\ref{udha}), and is also the reason that the \othree\
filter has been
placed ``downstream'' of the collimator. The manufacturer's
calculations for the residual reflection were consistent with
these specifications for normal incidence.

The science goals of the instrument led us to an initial choice of
narrow-band \othree\ filters centred at around 5000\AA, 5034\AA\ and
5058\AA.  The \FWHM\ of 31-35\AA\ is deliberately small to minimize the
background noise, and the filters are designed to be tiltable by up to
6\degree\ so that they can be tuned to the systemic velocity of each
target.  Optical tolerances on the 195~mm diameter circular filters were
held tight in order to achieve maximum homogeneity across the field and
at large tilt angles.  However, as the filters are rather near the pupil
for the reason mentioned in the previous paragraph, there is significant
shift in the bandpass over the field and this effect is amplified when
the filters are tilted, setting limits on the usable field size in the
direction of tilt. 

For the 5034/31.4\AA\ filter at
0\degree\ incidence the shift in bandpass over the field reduces
the useful bandpass, in the sense of unbiassed velocity coverage, to
about 29.6\AA. At 6\degree\ incidence this quantity is reduced to
17.0\AA, or 1000~\kms\ in velocity. The bandpass becomes even
smaller when the diagonals of the field are considered. Whether
the reduced bandpass is acceptable depends on parameters of the
object under study, most obviously its size.

\subsection{Mechanical design}
\label{md}

The mechanical design is in most respects straightforward, the main
priority being rigidity.  Stiffness is vital since relative displacement
of the two images by a few microns causes tens of~\kms\ shift in
velocity.  Structural analysis\footnote{JH, July 1999.}
predicts that the maximum flexure in either spectral or spatial directions
should correspond to an image displacement of only about 0.02 pixel
(0.01\arcsec\ on sky) when the spectrograph is moved through the full
range of positions encountered at either the Cassegrain or Nasmyth
focus.  This degree of rigidity
is more than sufficient since small movements of the
detector are likely to be considerably more significant. 

\begin{figure}

\includegraphics[width=12cm,angle=0]{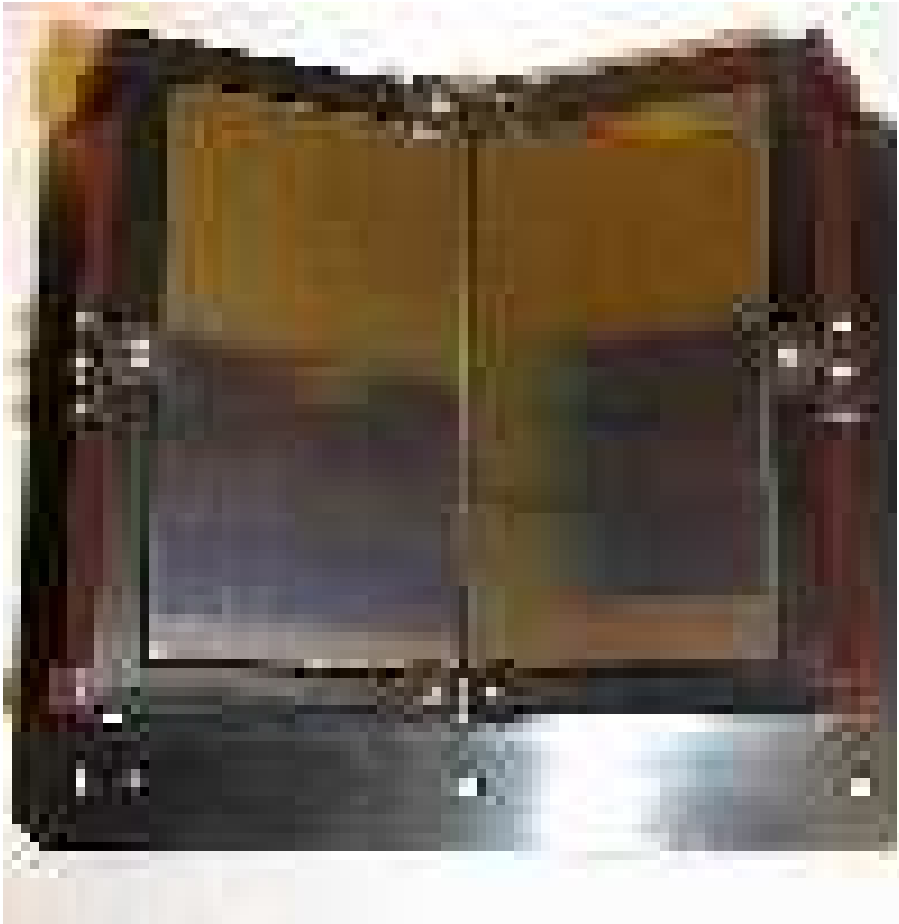}
 \caption{Photograph of the  unusual grating mount
in which two identical gratings (but with blaze direction
reversed) are oriented in a `concave' arrangement.}
\label{fig:gratings}
\end{figure}

A second problem to be addressed was that of butting 
together the gratings.  An
earlier ``convex'' arrangement in which the gratings would be
replicated on a prism-shaped substrate was rejected on the
grounds that the fabrication was too critical: once constructed,
no adjustment of the parallism of the two sets of grooves 
would have been possible, and of course 
the angle between the two grating normals would have been fixed also.
In the ``concave'' arrangement chosen (Fig.~\ref{fig:gratings})
the butted edges are bevelled at the appropriate angle to minimise light
loss at the join, and the two gratings held in place by the mount. 

\begin{figure}

\includegraphics[width=12cm,angle=0]{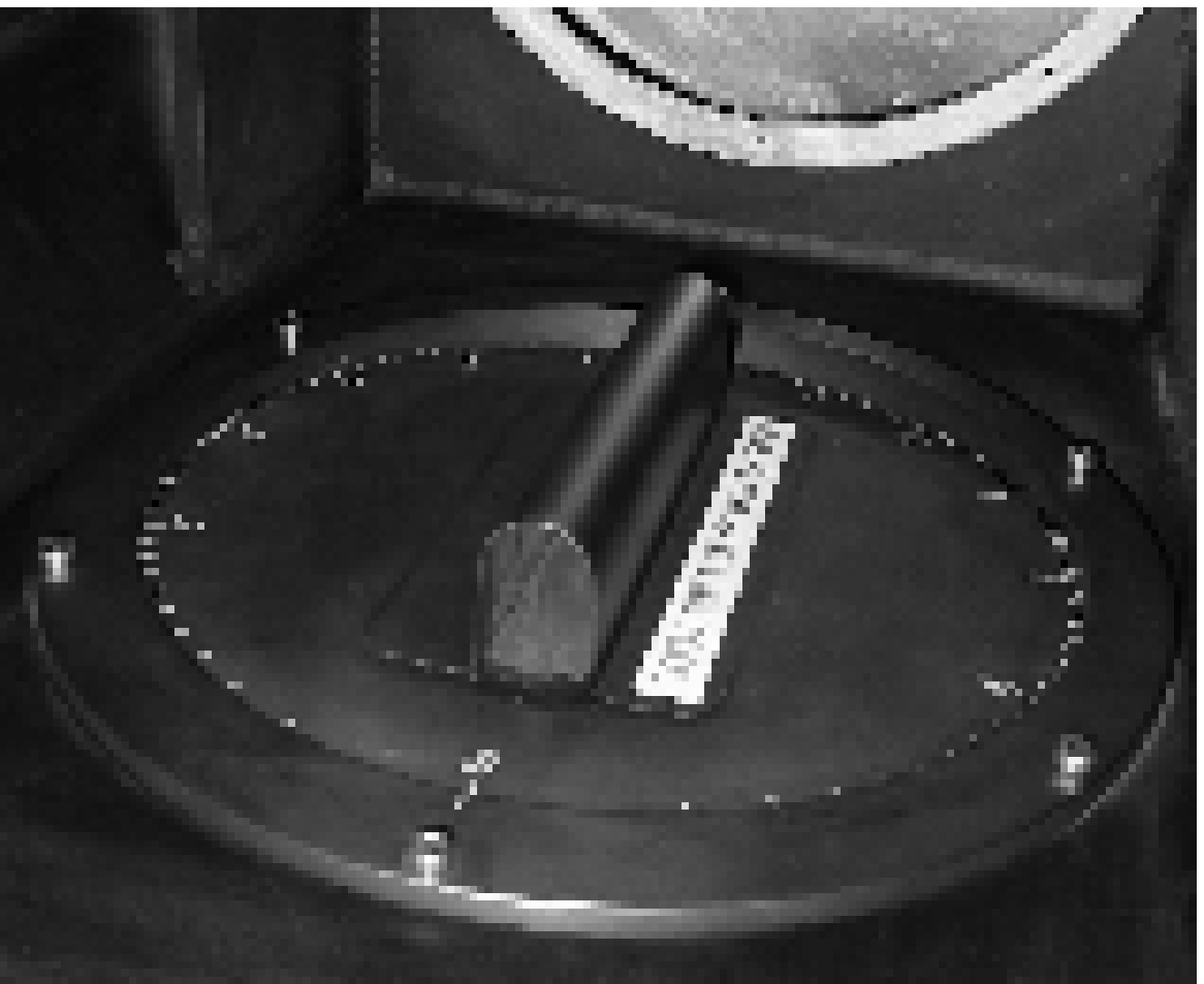}
 \caption{Photograph of the filter area, showing  the tool
used to tilt the filter by a calibrated amount
by altering the alignment of two wedged surfaces.
}
\label{fig:filter}
\end{figure}

The spectrograph incorporates just two (d.c.) motors, one to move the
calibration mask (Fig.~\ref{fig:mask}) into the
focal plane and one to drive a plate across the beam 
to serve as a shutter.  
Tilting of the \othree\ filter has to be done by hand, after opening
an access hatch. Tilting is achieved by rotating one wedged surface
against another, facilitated by a tool on which the corresponding tilt
angle is marked (Fig.~\ref{fig:filter}).  The operation requires the
telescope (in the case of the WHT) to be parked at the zenith and
takes about 5 minutes, including slewing.
A complete filter change requires considerably more time and targets have so
far been scheduled to avoid the neccessity of this being done
during the night.

\subsection{Calibration}
\label{calibration}
\begin{figure}

 \includegraphics[width=12cm,angle=0]{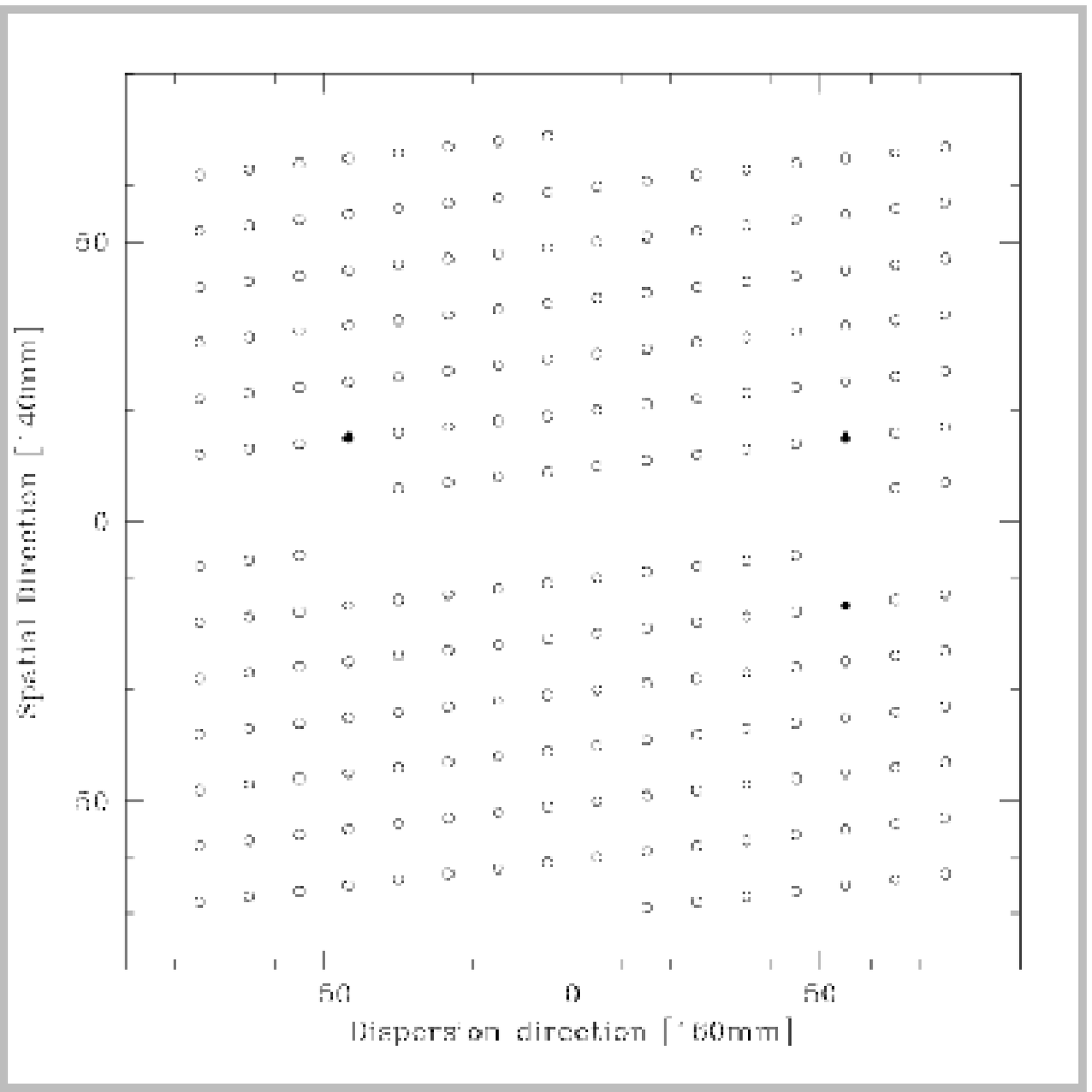}
 \caption{Focal plane calibration mask: 0.20~mm diameter holes 
in the metal mask are so
arranged that the resulting spectra (dispersed horizontally) do not overlap
(in order to determine the orientation of the images unambiguously
three holes are 50\% oversized).
 The large number of holes facilitates an over-constrained solution for
the wavelength-dependent mapping of the focal plane onto the detector. 
Illumination with a spectral lamp will give rise to 2-4 points per
hole, depending upon the filter in use.  
The single slit near the centre can be used when (e.g.) a stellar
spectrum is necessary for an independent velocity calibration.} \label{fig:mask}
 \end{figure}

As discussed in the references in \S\ref{rCDI},
calibration of CDI data requires taking account of
the internal distortions, which are dominated by the approximately
parabolic distortion resulting from the reflection at the
grating (``slit curvature''), which is an additive 
effect when one compares counter-dispersed image pairs. 
Other, smaller, effects include the discrepancies in the 
position of objects in the two images as a result of 
the inevitable differences between the optics in the two cameras.

A map of all the distortion terms, including wavelength dependences, can
be determined by the use of a calibration mask (Fig.~\ref{fig:mask})
which can be placed in the focal plane.  In combination with spectral
line lamps (typically CuAr and CuNe) and continuum lamps a
spectrum-dependent calibration over the whole field can be obtained. 
On-sky calibrations are only necessary to check photometry, though
Galactic \PN\ (with catalogued radial velocities) are regularly observed
to cross-check procedures.  Although we are still refining these
procedures the velocity residuals are currently no greater than 28~\kms\
(see \S\ref{pnso}). 

The final step in the calibration is to obtain accurate astrometry for
the newly-found \PN.  This is done by defining a cartesian coordinate
system in which the positions of a number of foreground stars are
defined by the coordinates about which the counter-dispersed spectra are
symmetric.  For example if the intensity-weighted average position of a
stellar spectrum in one image is $(x_L,y)$ and in the other $(x_R,y)$,
then $((x_L + x_R)/2,y)$ can be taken as the stellar position.  These
positions are then compared with astrometric catalogues to obtain a
plate solution in the usual way.  The sky co-ordinates of the \PN\ calculated on
this basis will be correct,  but the velocities will have
an offset, in this example, corresponding to the intensity-weighted mean
wavelength of the filter.  But this offset does not need to be
determined explicitly since the calibration mask fixes the absolute
wavelength scale for each pair of \PN\ detections.

\subsection{Project overview}
\label{overview}

The PN.Spectrograph consortium, whose members are listed on the
project website$^4$ was formed in 1997.
The project is funded by means of grants received
from the national funding agencies of the participating institutes, and
from ESO.  Total expenditure up to the commissioning of the instrument
was EUR~212k, which includes seven man-months of mechanical workshop
effort provided directly by the Dutch astronomy foundation ASTRON. 
Academic labour and travel costs were not charged to the project. 
Also, CCD detectors and cryostats are not included in the budget. 
By prior agreement the spectrograph makes use of the observatory
detectors and data management systems.

The optical design was contracted to Prime Optics (Australia) who worked
closely with the workshops of the RSAA where the mechanical design was
carried out. Contracts for the
construction of the camera optics and the collimator optics
went out to INAOE (Mexico) and to the RSAA optical workshop,
respectively, in March 1999.
Acceptance testing was carried out in May 2000 but this was inconclusive
and the optics could not be accepted until 
December, following further tests and corrective work by
RSAA.  The optics were then coated by the Australian company Rofin during
February-March 2001. 

In the meantime a Phase B study, to identify any remaining problems such
as stray light and flexure, had been carried out by RSAA and project
members.  Completed in July 1999, it showed that a double specular
reflection (zeroth order) between the gratings (which `face' each other in the
concave arrangement chosen - see Fig.~\ref{fig:ePNS}) would give rise to a
significant ghost image.  This problem was solved
by chosing different gratings, which were then ordered from the
Richardson Grating Laboratory. 

ASTRON constructed the instrument housing and many of the
component mountings, which  were shipped to Australia in July 2000 for
assembly.  
The first of the filters, from BARR associates, arrived in mid-May.

The RSAA workshops also manufactured the adapter flange for the WHT
Cassegrain (a different one is required for the TNG), the adapters for
the WHT cryostats, the tilt-tunable filter holder, the grating assembly
and the two motorized units (shutter and calibration mask).  Finally,
the RSAA was contracted to correct any residual mechanical errors, and
to integrate and align the instrument. 

During final testing, conducted by RSAA and Prime Optics staff,
a subtle scattered-light problem appeared which had not
been identified during the Phase B study. It again
involved multiple reflections
from the grating assembly and appeared when the (small)
reflectance of the CCD detectors was included. The problem
was solved by an additional baffle.

\subsection{The PN.S in operation}
\label{pnso}

The instrument was shipped to La Palma in the Canary Islands in June
2001, but unfortunately misdirected to Las Palmas, 
before arriving at the 
Observatory\footnote{Observatorio de la Roque de los Muchachos.} 
on July 10. 
The \ePNS\ was commissioned at the 4.2m WHT  of the
ING\footnote{Isaac Newton Group of telescopes.} on July 16, 2001
\citep{Merr01}.  It was then operated for a further three nights and has
since been used in two observing runs (September 2001 and March 2002). 
All observations were made with the EEV-12 and EEV-13 detectors, and to
date only the 5034\AA\ filter has been used, at either 0\degree\ tilt
($\lambda_c$ = 5034.3) or 6\degree\ tilt ($\lambda_c$ = 5027.1).

\begin{figure}

 \includegraphics[width=15cm,angle=0]{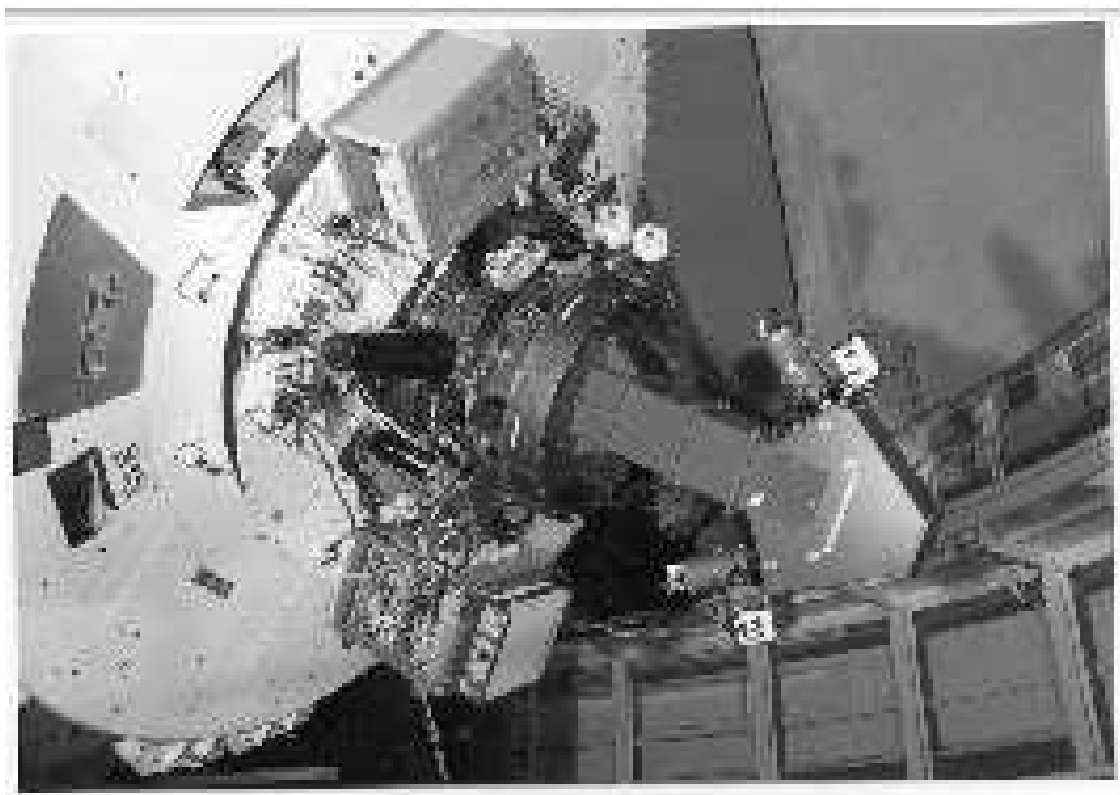}
\caption{The PN.Spectrograph at the WHT. 
}
\end{figure}

From measurements of galactic \PN\ with catalogued fluxes and of
spectrophotometric standard stars we were able to determine the system
efficiency (corrected for atmospheric losses) to be 33.3\%, in good
agreement with the design expectations (Table~\ref{table:calceff}).
Here the ``system efficiency'' includes telescope, instrument
including filter, and detector, and represents the throughput
at the peak of the filter profile. Since the dispersion has been
determined, it is easy to establish the peak throughput per \AA ngstrom
by measuring the flux in a slice through the peak of the
spectrum of a standard star and comparing this with the expected flux.
 
To use this information to establish a magnitude limit, we refer to
\S\ref{obs} from which we can see that at the distance of the
Virgo cluster  (15 Mpc) the
brightest \PN\ have $m^{*} \sim 26.6$.
At the above efficiency, in a one-hour observation with 1\arcsec\ seeing,
this corresponds to
about 1200 detected photons and a \SN\ of about 10.1 in each
arm of the spectrograph, using Equation~\ref{sn:cdi}. \PN\ which are 1.4
magnitudes fainter reach $S/N \sim 10 $ in 12.0 hours. Even at 
a distance of 25~Mpc ($m^{*} \sim 27.7$),
significant sampling of the PNLF can be achieved ($S/N \sim 7 $ for
$m^{*}+0.6$ mag in 14.6 hours).

Preliminary measurements show a differential displacement between the
left and right images of up to 3 pixels ($\sim 40$\micron) when the
telescope is rapidly  slewed through the full range of azimuth and
elevation, which is much more than that expected from flexure
calculations (\S\ref{md}).  Most of this effect seems to
be due to residual movements of the CCD detectors within the cryostats,
which was not included in the calculation. 

Fortunately, by taking arc lamp images of the focal plane mask at regular intervals
during observations, we are able to ensure that the velocity calibration
is not affected by flexure.  After slewing to each new target the
zero point of the flexure is determined by registration of the
calibration mask in the two cameras. Measurements have also shown
that the
displacements in either image are negligibly small during  normal guiding
($\pm 0.5$ pixel r.m.s.) so
the imaging is not compromised. 

\begin{figure}
 \includegraphics[width=12cm,angle=0]{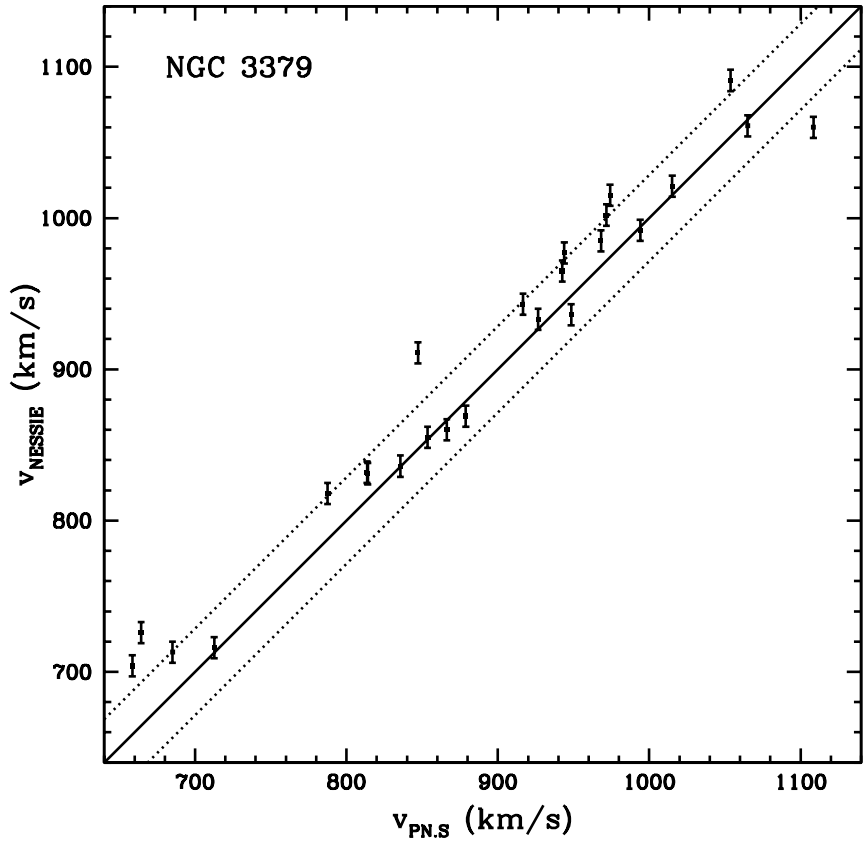}
 \caption{
Velocities of 25 planetary nebulae in NGC 3379, as measured with the PN.S
and NESSIE.  The solid line indicates a perfect correspondence, and the
dotted lines show uncertainty boundaries of $\pm$ 28 km~s$^{-1}$ for the
PN.S. The error bars show the NESSIE measurement uncertainties of 7
km~s$^{-1}$.
} \label{fig:velcalib}
 \end{figure}

The wavelength calibration, still being optimised, has tested out quite
satisfactorily during observations of objects with known velocities.  For
example, a comparison of 25 of our measured PN velocities in NGC 3379 ($D$
= 10 Mpc) with measurements from the NESSIE multi-fibre spectrograph on
the KPNO 4-m telescope \citep{cjd93} shows good
agreement, indicating our measurement uncertainty at this stage to be 28
km~s$^{-1}$ (Fig.~\ref{fig:velcalib}).

For conditions under which the seeing was worse than $\sim 0.8$\arcsec\
the image quality was uniform over the field.  Under better conditions,
it becomes evident that the image quality is slightly poorer towards the
corners of the field, as expected (see Fig.~\ref{fig:spots}).  Under very
good conditions the imaging quality is limited by the spectrograph
optics to $\sim 0.7$\arcsec. 

\begin{figure}
 \includegraphics[width=8cm,angle=0]{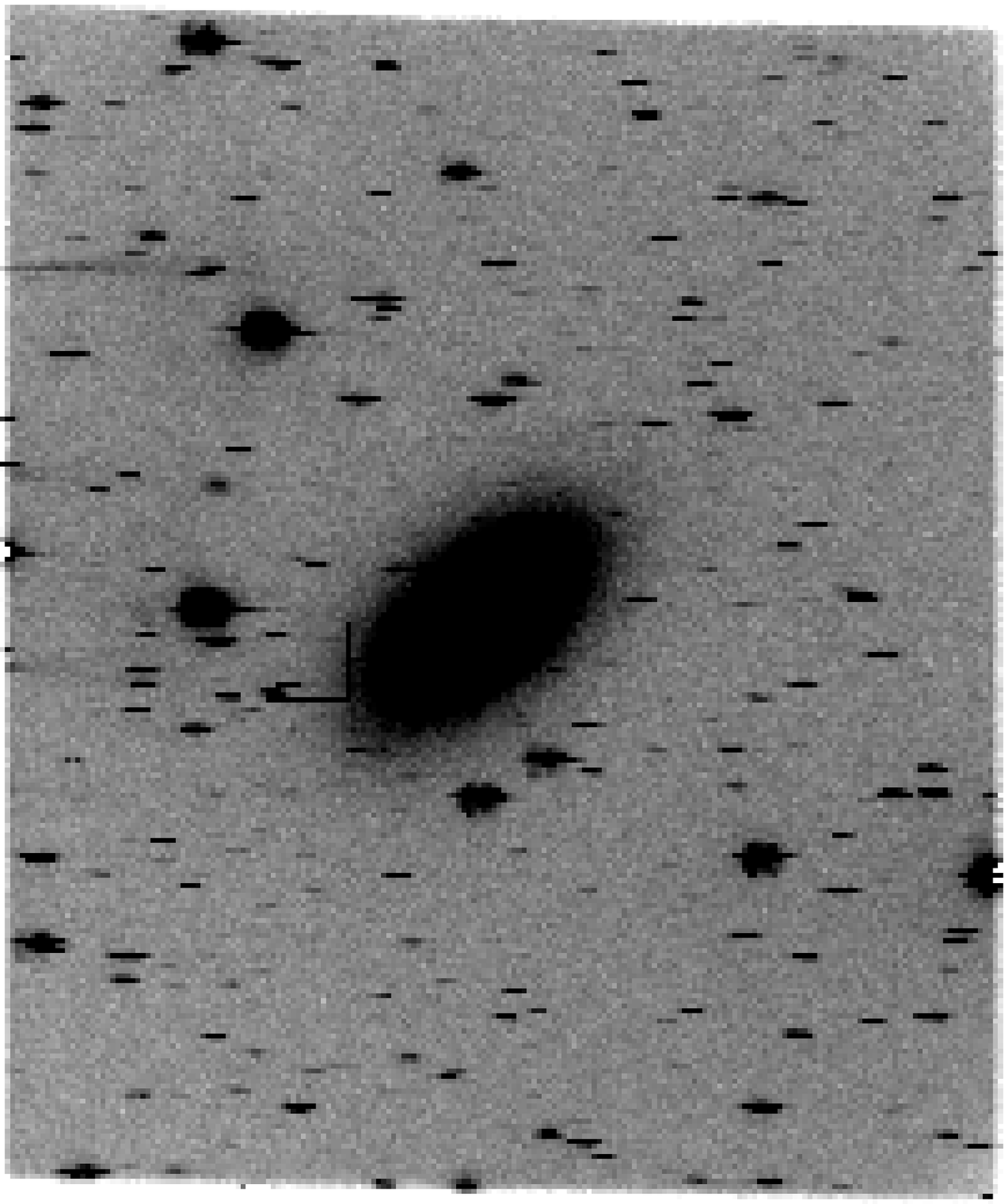} 
 \includegraphics[width=8cm,angle=0]{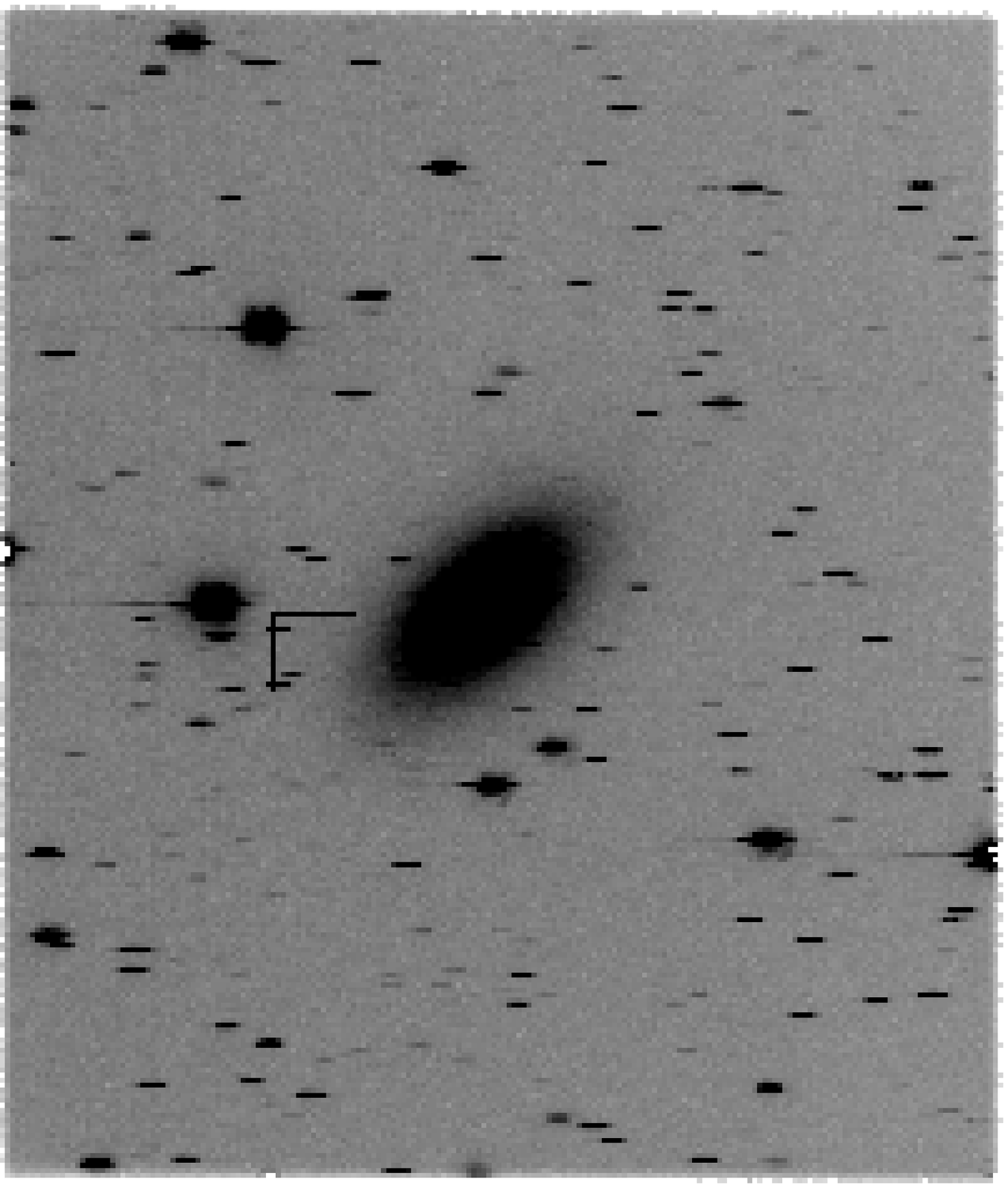}
 \vskip 0.1mm
 \includegraphics[width=8cm,angle=0]{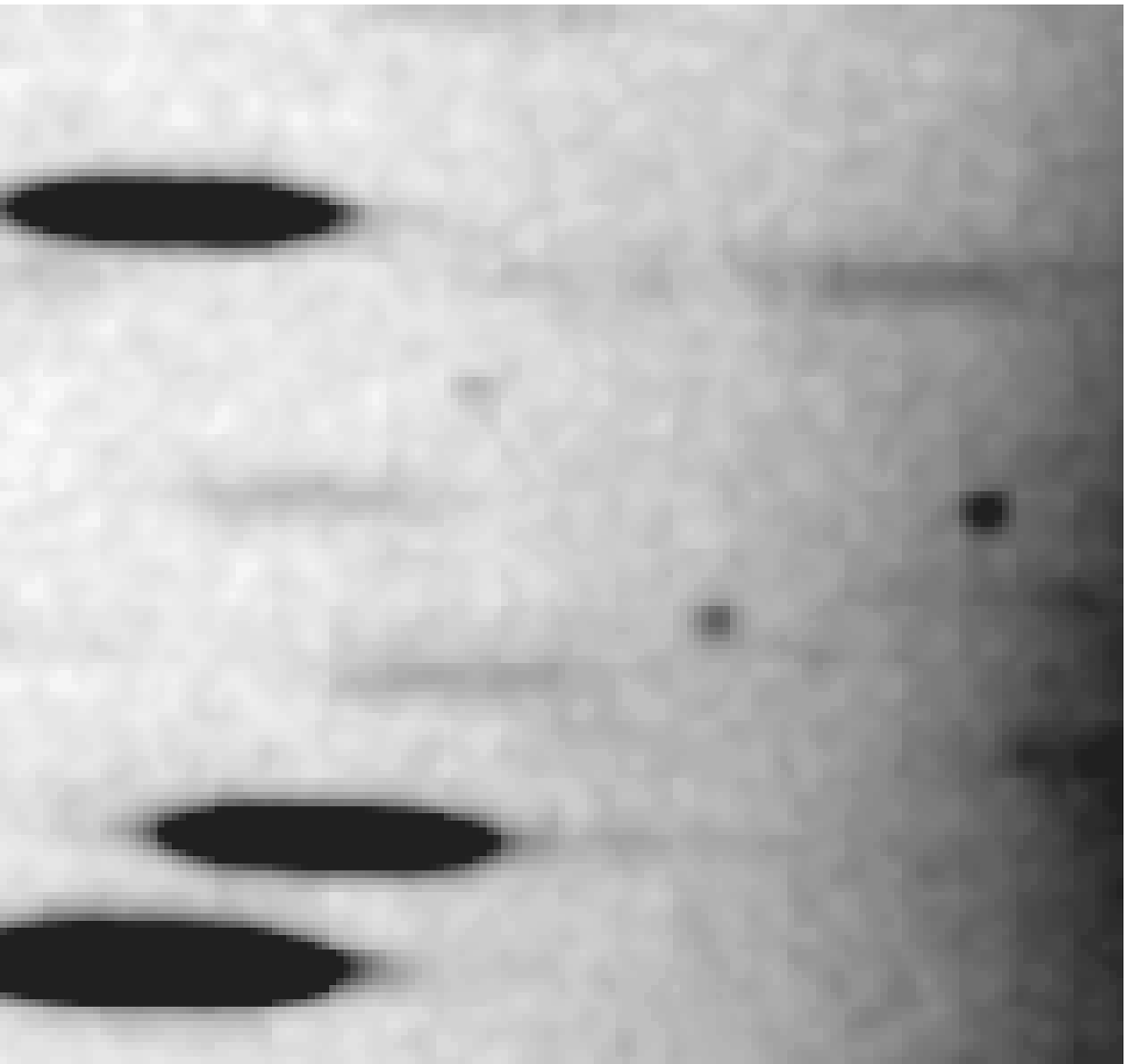} 
 \includegraphics[width=8cm,angle=0]{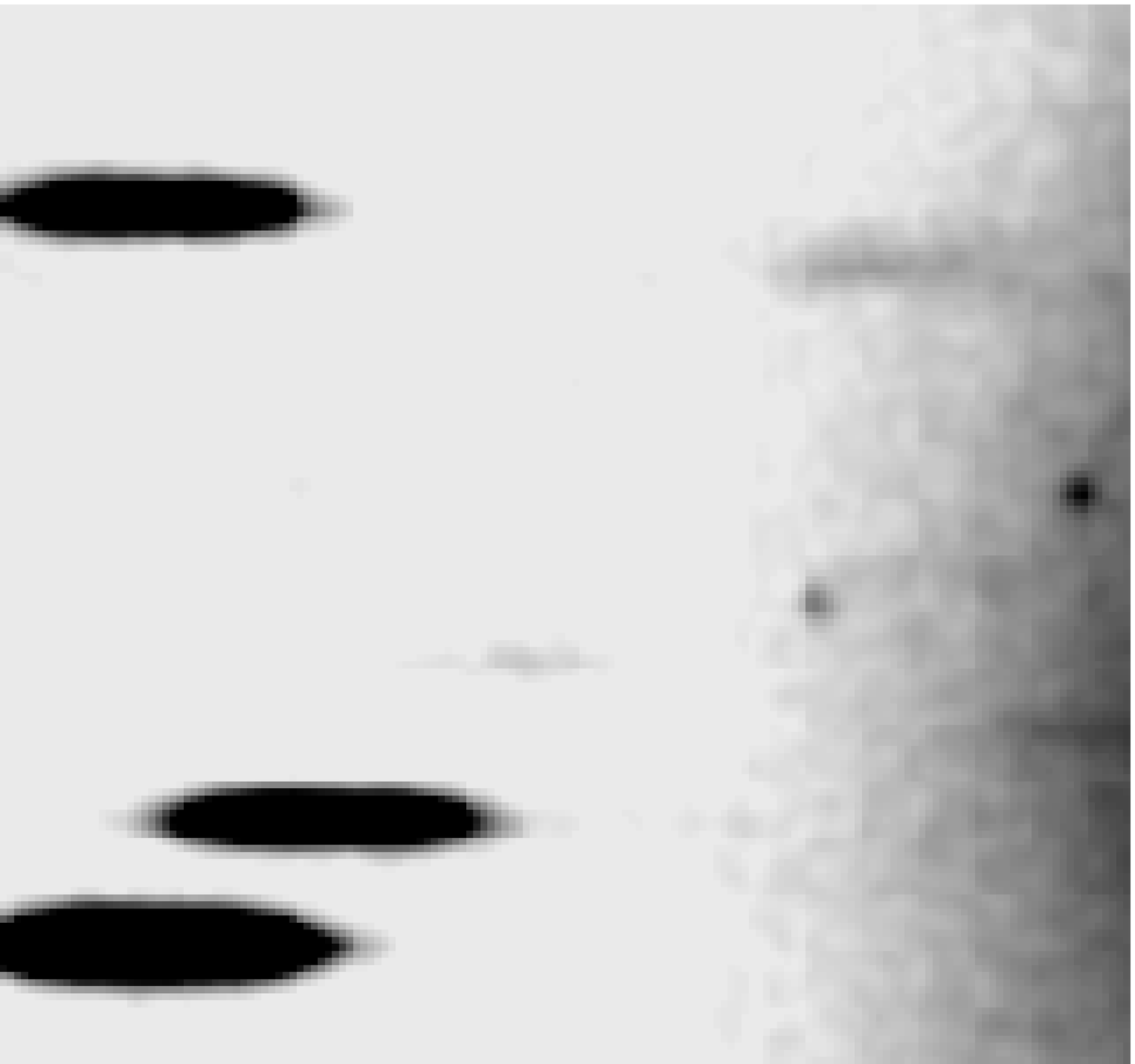} 

 \caption{Images of NGC~7457 taken with the PN.S in September
2001. The CDI pair is shown at the top; the lower panels show
detail in the boxed region, showing stellar `images' and the
displaced
positions of \PN. }
 \label{fig:someimages}
 \end{figure}

Images from one of the first observing sessions at the WHT 
are shown in Fig.~\ref{fig:someimages}.  The left- and
right-arm images appear nearly identical, since they are visually
dominated by the short, symmetric spectra of stars and by the broad,
integrated light of the galaxy (NGC~7457).  However the positions of the
\PN, some of which can be seen as point-like sources in the zoomed-in
images, are displaced between left and right images, as can be seen by
comparing them with the stellar features (cf Fig.~\ref{fig:pnseek1e}).

\section{Conclusion}
\label{conclusion} 
 We have constructed an instrument which can efficiently detect
extragalactic \PN\ and measure their velocities.  Using a novel
application of slitless spectroscopy, the data can be obtained in a
single observation with a single instrument, rather than the more
complex procedures previously necessary.  We have presented design
information for the highly optimised, dedicated instrument now in use at
La Palma, and first results have been presented.  In effect
the telescope efficiency can be nearly doubled by
this approach, opening up the prospect of routine observations of galaxy
kinematics at a distance of up to 25~Mpc with 4m-class telescopes. The
technique is applicable to larger telescopes.

\section{Acknowledgements}

The \ePNS\ would not have seen the green light of day without the generous 
financial support of: the Australian Research Council,
the European Southern Observatory, the Italian C.N.A.A. (Consiglio Nazionale per
l'Astronomia e l'Astrofisica), the Dutch N.W.O. (Nederlandse Organisatie voor 
Wetenschappelijk Onderzoek), the Observatory of Capodimonte and in  Britain the
Royal Society and  PPARC.

We thank the ING (operating the William Herschel Telescope on La Palma)
for their enthusiastic support, the Mt Stromlo workshops of the R.S.S.A. 
for their untiring efforts, and the Anglo-Australian Observatory.  We
also wish to thank Steve Rawlings, who generously assisted by swapping
his telescope time allocation after delays caused planning problems for
the project, and Rob Hammerschlag and and Felix Bettonvil for technical advice.

\section{Appendix A: Undispersed H$\alpha$ arm}
\label{udha}

An \Halpha\ imaging camera was included as a future option  in the
\ePNS\ design.  A dichroic can be placed after the last element
of the collimator, bringing an undispersed image to a third CCD
detector via a \Halpha\ filter and camera. We are now seeking to
fund this option. 
The motivation is to improve the detection efficiency of \PN\ and to
permit deep Halpha imaging of galaxies which we are already surveying for 
\PN.

For those \PN\ which are detected in \Halpha, the \Halpha\ image
constitute a cross-check of the velocity measurement. The image will also
serve as a check on the astrometric procedures used on the CDI
images, and as an indicator of contaminating objects.  For
example, the absence of an \Halpha\ counterpart to an \othree\
detection will indicate a background object instead of a PN: 
\Lyalpha\ emitters redshifted to the \othree\ band are detected
in significant numbers in
\othree\ surveys (e.g. \citet{Free99}).  Objects which are
relatively bright in \Halpha\ may be unresolved \Htwo\ regions.
Although the \ntwo\ $\lambda 6584$\AA\ line will often be
included in the \Halpha\ image, some information on the
\Halpha/\othree\ ratio would be obtained. 

%

%

While elliptical galaxies harbour much less cold and warm gas
than spirals, with deep enough exposures they mostly show
extended \Halpha\ emission \citep{Mac96}).  With our long
integration times (typically 10 hours on each galaxy, and with
an optical efficiency higher than that obtained with multi-purpose instruments), 
we expect to achieve some of the deepest
\Halpha\ images yet (for example, about a factor of 6 deeper than
the images of Macchetto et al.\ (op. cit.) 
in their survey at the ESO 3.6m
and 3.5m NTT).  These new emission maps may provide
useful insight into the ionization processes in elliptical
galaxies. Similar deep \Halpha\ images can be acquired in the
outer regions of S0 galaxies, where any remaining interstellar
gas is probably ionized by the intergalactic UV radiation. Long
integration times are required to detect this diffuse ionized gas
(see \citet{bfq97}), which may help to constrain the ionizing flux
and gas density \citep{malb99}.

\section{Appendix B: Rejected design solutions}
\label{rds}

Several other possibilities exist for the construction of a simultaneous
CDI instrument such as the \ePNS.  For future reference we briefly
summarise some that we have investigated. 

\subsection{Amplitude splitting}
\label{as}

\begin{figure}
\includegraphics[width=12cm,angle=0]{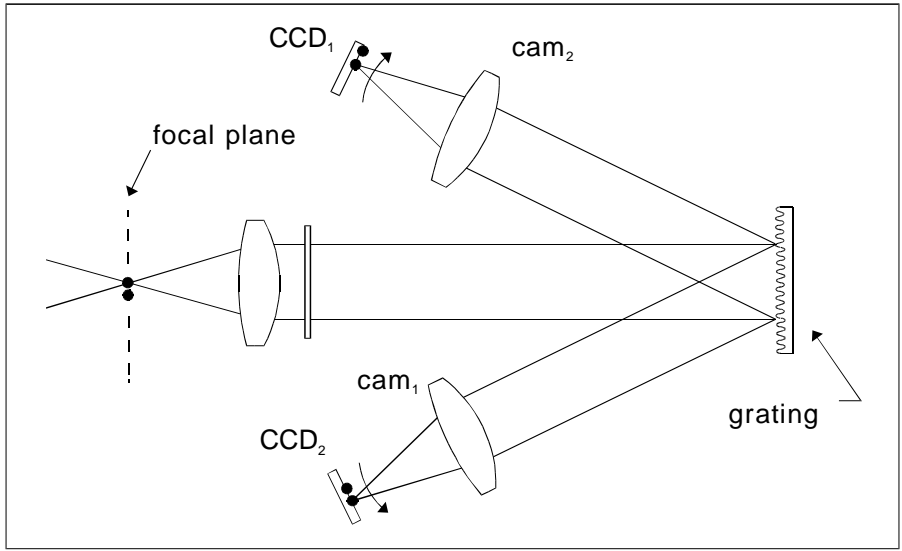}

\caption{Amplitude splitting of the wavefront.
}
\label{fig:amplitude.spl}
\end{figure}

Instead of a pair of gratings butted together (pupil splitting)
we consider a single grating illuminated at normal incidence 
(Fig.~\ref{fig:amplitude.spl}).
By choice of
grating constant, only the orders $m =$  -1, 0 and 1
are propagating.
Therefore, the only possible source of stray
light is from the $m=0$ order, which in this case would not be a great 
problem.
This is an elegant construction since
symmetry is guaranteed, there is no join
between the two gratings, and the mechanical mounting is trivial.

\citet{connes59} explored this approach in various
interferometer designs, and believed on the grounds of a scalar
analysis that a grating with a suitable symmetric  triangular
profile would split light at normal incidence into two beams with
nearly 50\% efficiency in each. Using electromagnetic code
we discovered that for a 1800\lmm\ grating the optimum efficiency 
(averaged over both  polarisations) is only 
64\% (in aluminium) at a blaze angle of
34\degree, slightly displaced with respect to the 32\degree\
expected on the basis of geometric.  This is well short of the nearly
100\% optical efficiency expected by Connes, and the difficulty
of construction does not justify its use when compared with a
holographically produced grating. Incidentally Connes'
expectation does work out correctly for the S-polarised
component.

We performed an efficiency optimisation for a holographic
grating, assuming a sinusoidal profile again in aluminium.  A
regime of high ($>80$\%) efficiency is found near 1250\lmm\ and
a groove depth of 0.165~\micron\ (the p- and s-polarisation
efficiencies are about equal). 
The cost of a custom-made grating would have been prohibitive though,
while the highest efficiency for an industry standard 
holographic grating was found to be
around 66\% in unpolarised
light, with 1800\lmm\ and a groove depth of
0.16~\micron\footnote{This  calculation, using the {\tt gtgr4}
code mentioned earlier, was kindly
confirmed by Jobin-Yvon, a French manufacturer.}. 

\subsection{Polarization splitting}
\label{pols}

The problem of obtaining high efficiency from a grating is made
much easier if  the grating is illuminated by light of only one polarization. 
A CDI instrument arises if the light is split into two
linearly-polarized beams, each of which is fed to a separate
grating
{Fig.~\ref{fig:polarization.spl}). The idea calls for a more
complicated mechanical design though, and the resulting
asymmetries were feared to lead to flexure problems.

\begin{figure}
\includegraphics[width=7cm,angle=0]{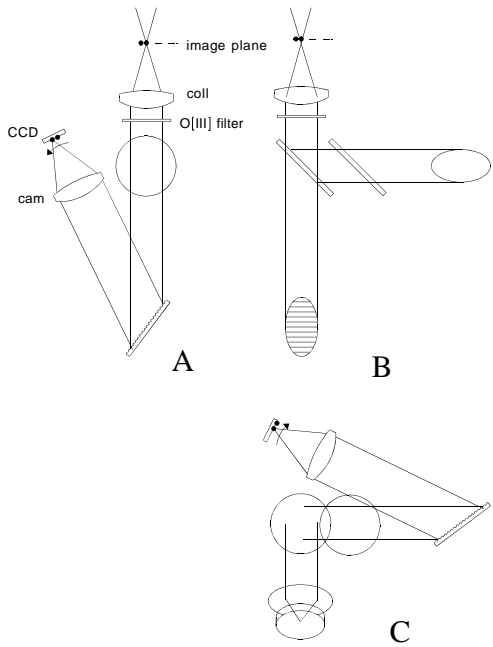}
\caption{CDI using polarisation splitting. The unpolarised light
is  divided into two beams by means of a polarising beamsplitter
(see view B, in which the reflected beam also traverses a
compensation plate). The reflected light (view C, looking down
from the telescope axis) and  the transmitted light (view A)
encounter similar, but polarisation-optimised, spectrographs.
}
\label{fig:polarization.spl}
\end{figure}

\subsection{Transmissive gratings}
\label{tg}

The transmissive equivalent of pupil splitting would correspond
to two grisms, side-by-side.  This would make it 
possible to use  one  camera instead of two, the two
images falling side-by-side on a single large format CCD. During
our study we were unable to find a grism with sufficient dispersion and
efficiency. 

We did investigate the idea of a transmissive phase mask, in which the
collimated beam is passed through a crenellated dielectric structure
which causes it to diffract into pairs of orders (similar to the
case in \S\ref{as}). Provided the zeroth
order is minimised by selecting the groove depth, and higher orders are
cut off, high efficiency can be achieved in the --1 and +1 orders.  A
review is given by \citet{walker}, who show the effect
of varying all parameters including the duty cycle of the structure,
indicating that high efficiency can be obtained. 

In an early (1994) scaled experiment using a
microlithographically produced mask and UV light,
we verified that high transmitted efficiencies ($\sim 75$\%) can 
be reached\footnote{This was performed by
NGD at the University of Sydney, Australia, assisted by Dr P.
Krug.}  At the time, it was difficult to obtain these structures
in large sizes.  Since then, the closely related 
VPH (volume phase holographic) gratings have
become available, and transmissive gratings may be a more viable
option. 

\section{Appendix C: Accuracy of PSF-fitting photometry and positions}
\label{centr}

Here we calculate the accuracy of object centroids derived by means of
PSF fitting. 

Let $f_i$ be the data: intensities on the pixels $i$ at positions
$x_i,y_i$ (in units of pixels) on the image plane. We try to model these data as 
$ A \PSF(x_i-\mu_x,y_i-\mu_y) \Delta^2$, where $\Delta$ is the pixel
width, and $\PSF$ is the PSF normalized to total intensity 1. 
$A$ is the intensity of the star, and $(\mu_x,\mu_y)$ are the position
of the star, to be fitted for. 
Write 
\[
\chi^2=\sum_i \left(f_i - A \PSF(x_i-\mu_x,y_i-\mu_y) \Delta^2\right)^2
                /\sigma_i^2
\]
where $\sigma_i$ is the error on the measured intensity in pixel $i$.
Then the minimum of $\chi^2$ gives the best-fit PSF, and the second
partial derivatives of $\chi^2$ with respect to any two parameters
give twice the inverse covariance matrix. In what follows we ignore
covariances between the different variables $A$, $\mu_x$ and $\mu_y$,
as is appropriate for bisymmetric PSF's. 

At the best-fit value (we may assume without loss of generality that
$\mu_x=\mu_y=0$) we obtain the inverse variance on the star's intensity
$A$ as 
\[
\hbox{Var}(A)^{-1} =
0.5\ \partial^2\chi^2/\partial A^2=\sum_i \PSF (x_i,y_i)^2 \Delta^4 /\sigma_i^2
\]
which reduces to 
\[
                 =\left(\int\int \PSF^2\ dx dy\right) \Delta^2 / \sigma^2 
\] 
if the PSF is fully sampled and $\sigma_i$ constant (i.e.,
background-limited data). All integrals are over the range
$0 - \infty$.

The inverse variance of the best-fit position is similarly (removing
terms which go to zero at the best fit)

\[
\hbox{Var}(\mu_x)^{-1} =
0.5\ \partial^2\chi^2/\partial\mu_x^2 =
\sum_i A^2 (\partial\PSF/\partial x)^2 \Delta^4 /\sigma^2
  =\left(\int\int (\partial\PSF/\partial x)^2 dx\ dy\right) A^2 \Delta^2 / \sigma^2
\]
(and similarly for $\mu_y$).
These two relations can be combined to give
\[
         \delta \mu_x={\delta A\over A} \times \sqrt{\int\int \PSF^2\ dx dy \over
\int\int (\partial\PSF/\partial x)^2\ dx dy}
\]
where $\delta A = \hbox{Var}(A)^{1/2}$ is the 1-$\sigma$ error on
$A$, etc.

The centroid error therefore depends on the significance $A/\delta A$
of the detection of the star, and on a geometric factor governed only
by the shape of the PSF.

For a gaussian PSF, dispersion $s$, we find 
\[
\hbox{Var}(A)^{-1}=\Delta^2/(4 \pi s^2 \sigma^2)
\]
hence 
\[
\delta A    =\sqrt{4 \pi} s \sigma /\Delta
\]
and
\[
\hbox{Var}(\mu_x)^{-1}=A^2 \Delta^2 /(8 \pi s^4 \sigma^2)
\]
hence 
\[
\delta \mu_x    =\sqrt{8 \pi} s^2 \sigma / (\Delta A)
                    =\sqrt2 s {\delta A\over A}
\]

For a Moffat function PSF of the form 
\[
\PSF={\beta-1\over \pi a^2} \left(1+{r^2\over a^2}\right)^{-\beta}
\]
we get
\[
      \delta \mu_x=a {\delta A\over A} \sqrt{2 \beta+1\over \beta (2\beta-1)} 
\] 
If we write $a$ as $\frac12 \hbox{FWHM}/\sqrt{2^{1/\beta}-1}$, 
we find that for $\beta> 1.5$, which covers PSF's with tails as
shallow as $r^{-3}$
\[
         \delta \mu_x={\delta A\over A} \times 0.67 \times \hbox{FWHM} \pm
10\%.
\]         
(The gaussian is the limit $\beta\to\infty$; in this case the
coefficient is 0.6.)

\end{document}